\documentclass[aps,prl,twocolumn,superscriptaddress,preprintnumbers,amsmath,amssymb,showkeys,floatfix,nofootinbib,reprint]{revtex4-2} 

\usepackage{graphicx}
\usepackage{dcolumn}
\usepackage{bm}
\usepackage{color}
\usepackage{txfonts}
\usepackage{microtype}
\usepackage[english]{babel}
\usepackage{slashed}
\usepackage{gensymb}
\usepackage{epsfig}
\usepackage {appendix}
\usepackage{multirow}
\usepackage{array}
\usepackage{amsmath}
\usepackage{booktabs} 
\usepackage{makecell} 
\usepackage[T1]{fontenc}
\usepackage{lmodern}
\usepackage{graphicx}
\usepackage[utf8]{inputenc}
\usepackage{hyperref}
\usepackage[utf8]{inputenc}
\usepackage{newunicodechar}
\newunicodechar{，}{,}
\DeclareUnicodeCharacter{2212}{-}
\newcommand{\tabincell}[2]{\begin{tabular}{@{}#1@{}}#2\end{tabular}}
\begin{document}
	
	\renewcommand{\figurename}{FIG}	
	
	\title{Nuclear Ptychoscopy: A Ptychographic Framework for Nuclear Spectroscopy}
	
	\author{Ziyang Yuan}
	\affiliation{Key Laboratory of Nuclear Physics and Ion-Beam Application (MOE), Institute of Modern Physics, Fudan University, 200433 Shanghai, China}
	\affiliation{Research Center for Theoretical Nuclear Physics, NSFC and Fudan University, 200438 Shanghai, China}
	
	\author{Yifei Zhang}
	\affiliation{Key Laboratory of Nuclear Physics and Ion-Beam Application (MOE), Institute of Modern Physics, Fudan University, 200433 Shanghai, China}
	\affiliation{Research Center for Theoretical Nuclear Physics, NSFC and Fudan University, 200438 Shanghai, China}
	
	\author{Yonggong Teng}
	\affiliation{Key Laboratory of Nuclear Physics and Ion-Beam Application (MOE), Institute of Modern Physics, Fudan University, 200433 Shanghai, China}
	\affiliation{Research Center for Theoretical Nuclear Physics, NSFC and Fudan University, 200438 Shanghai, China}
	\affiliation{School of Data Science, Fudan University, Shanghai 200433, China}
	
	\author{Hongxia Wang}
	\affiliation{Kaiyuan International Mathematical Sciences Institute, Changsha, 410083, China}
	
	\author{Fengjiao Gan}
	\affiliation{The School of Communication Engineering, Hangzhou Dianzi University, Hangzhou, 310005, China}
	
	\author{Hao Wu}
	\affiliation{The School of Communication Engineering, Hangzhou Dianzi University, Hangzhou, 310005, China}
	
	\author{Xinchao Huang}
	\affiliation{Hefei National Laboratory for Physical Sciences at Microscale and Department of Modern Physics, University of Science and Technology of China, Hefei, Anhui 230026, China}
	
	\author{Tianjun Li}
	\affiliation{Hefei National Laboratory for Physical Sciences at Microscale and Department of Modern Physics, University of Science and Technology of China, Hefei, Anhui 230026, China}

	\author{Ziru Ma}
	\affiliation{Hefei National Laboratory for Physical Sciences at Microscale and Department of Modern Physics, University of Science and Technology of China, Hefei, Anhui 230026, China}
	\author{Linfan Zhu}
	\affiliation{Hefei National Laboratory for Physical Sciences at Microscale and Department of Modern Physics, University of Science and Technology of China, Hefei, Anhui 230026, China}
	
	\author{Zhiwei Li}
	\affiliation{Key Laboratory of Magnetism and Magnetic Functional Materials (Lanzhou University)，Ministry of Education, Lanzhou 730000, China
	}
	
	\author{Wei Xu}
	\affiliation{Beijing Synchrotron Radiation Facility, Institute of High Energy Physics, Chinese Academy of Sciences, Beijing 100049, China}
	
	\author{Yujun Zhang}
	\affiliation{Beijing Synchrotron Radiation Facility, Institute of High Energy Physics, Chinese Academy of Sciences, Beijing 100049, China}
	
	\author{Ryo Masuda}
	\affiliation{Faculty of Science and Technology, Hirosaki University, Bunkyo-cho, Hirosaki-shi, Aomori 036-8561 Japan}
	
	\author{Nobumoto Nagasawa}
	\affiliation{Precision Spectroscopy Division, Japan Synchrotron Radiation Research Institute, Sayo, Hyogo 679-5198, Japan}
	
	\author{Yoshitaka Yoda}
	\affiliation{Precision Spectroscopy Division, Japan Synchrotron Radiation Research Institute, Sayo, Hyogo 679-5198, Japan}
	
	\author{Jianmin \surname{Yuan}}
	\affiliation{Institute of Atomic and Molecular Physics, Jilin University, Changchun 130012, China}
	
	
	\author{Xiangjin Kong} 
	\email{kongxiangjin@fudan.edu.cn}
	\affiliation{Key Laboratory of Nuclear Physics and Ion-Beam Application (MOE), Institute of Modern Physics, Fudan University, 200433 Shanghai, China}
	\affiliation{Research Center for Theoretical Nuclear Physics, NSFC and Fudan University, 200438 Shanghai, China}
	
	\author{Yu-Gang Ma}
	\email{mayugang@fudan.edu.cn}
	\affiliation{Key Laboratory of Nuclear Physics and Ion-Beam Application (MOE), Institute of Modern Physics, Fudan University, 200433 Shanghai, China}
	\affiliation{Research Center for Theoretical Nuclear Physics, NSFC and Fudan University, 200438 Shanghai, China}
	\date{\today}

	\date{\today}
	
	\begin{abstract}
		Accessing both amplitude and phase of nuclear response functions is central to fully characterizing light–matter interactions in the X-ray-nuclear regime. Recent work has demonstrated phase retrieval in two-dimensional time- and energy-resolved spectra, establishing the feasibility of phase-sensitive nuclear spectroscopy. Here, we introduce Nuclear Ptychoscopy, a ptychographic framework that adapts algorithms from coherent diffractive imaging to nuclear spectroscopy, enabling reconstruction of the complex response function by exploiting redundancy in two-dimensional spectra. We develop three complementary reconstruction schemes tailored to distinct experimental scenarios: reconstruction with a known analyzer response, blind reconstruction, and reconstruction incorporating partial prior information. In parallel, we develop geometric analysis techniques that elucidate algorithmic behavior and contribute new tools to ptychography. The framework is validated through experimental data and simulations, demonstrating its versatility across diverse nuclear spectroscopy scenarios and bridging nuclear spectroscopy with ptychography. Beyond advancing quantitative nuclear spectroscopy, our framework opens new opportunities for metrology, coherent control, and quantum applications in the X-ray–nuclear regime.   	
	\end{abstract}
	
	\maketitle
	
	\section{Introduction}
	
	
	Ptychography is a powerful computational imaging technique that reconstructs both the amplitude and phase of a wavefield from a series of overlapping measurements \cite{miao2025computational,pfeiffer2018x}. By exploiting redundancy in the measured data, it enables high-fidelity, model-independent retrieval of information that is otherwise inaccessible in intensity-only measurements \cite{faulkner2004movable,rodenburg2004phase,rodenburg2007hard,thibault2009probe}. Its success in coherent X-ray imaging \cite{chapman2010coherent}, electron microscopy \cite{thibault2008high}, and optical metrology \cite{claus2013dual} demonstrates how relative shifts and overlaps in data can be leveraged to reconstruct complex-valued fields with remarkable performance. 
	
	
	A similar need arises in nuclear spectroscopy~\cite{greenwood2012mossbauer,yoshida2021modern}. M\"{o}ssbauer nuclei, with their exceptionally narrow transitions and high-quality factors, serve as sensitive probes for metrology and quantum optics in the X-ray regime~\cite{PhysRevLett.77.3232,rohlsberger2005accelerating,rohlsberger2010collective,rohlsberger2012electromagnetically,heeg2013vacuum,vagizov2014coherent,Fano2015,heeg2015tunable,haber2016,haber2017rabi,heeg2021coherent,Chumakov2018,larss,Shvydko2023,r2hf-9qn9}. Conventional techniques—such as employing Synchrotron M\"{o}ssbauer Source~\cite{potapkin201257fe,kupenko2024nuclear,mitsui2018variable,fujiwara2024synchrotron} or time-integrated spectroscopy with a Doppler-driven analyzer~\cite{coussement1996time,l2000experimental,callens2002stroboscopic,callens2003principles}—record only intensities, discarding crucial phase information. While these methods provide energy spectra, they cannot reconstruct the full complex response of the nuclear system. Recent developments have shown that both amplitude and phase can be extracted from two-dimensional (2D) time- and energy-resolved spectra~\cite{Yuan2025,negi2025energy,herkommer2020phase,wolff2023unraveling}.  These works demonstrate the feasibility of phase-sensitive nuclear spectroscopy, while highlighting the need for a systematic and versatile framework to reconstruct nuclear response functions across a wider range of experimental conditions.
	
	\begin{figure*}[ht]
		\setlength{\abovecaptionskip}{-0.2cm} 
		\centerline{\includegraphics[width=1.00\linewidth]{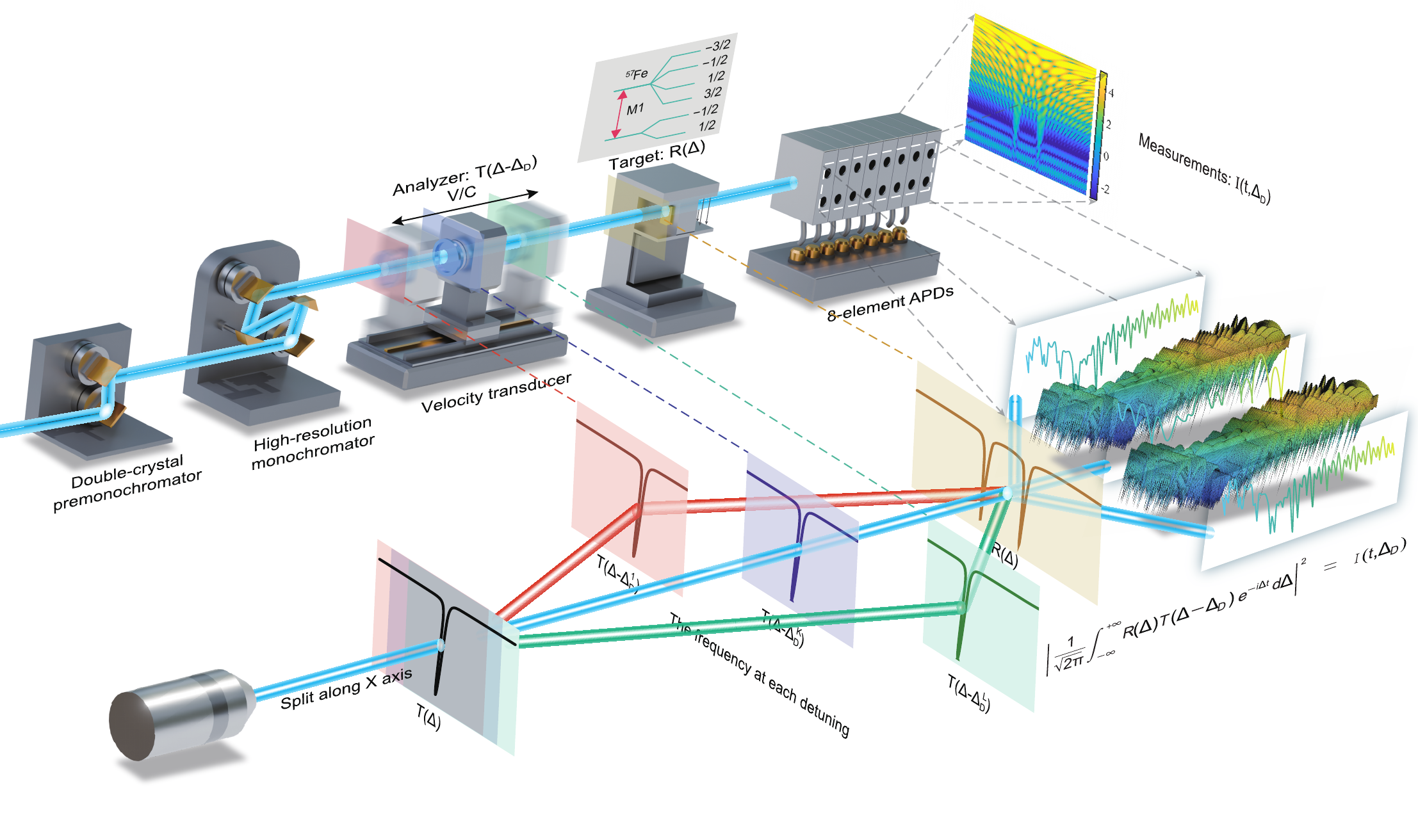}}
		\centering
		\begin{picture}(10,5)
		\end{picture}
		\caption{Experimental setup and its analogy to ptychography. Synchrotron radiation first passes through the monochromators, then through an analyzer with transmission function $T(\Delta-\Delta_D)$ mounted on a M\"{o}ssbauer drive, and is subsequently scattered by a $^{57}$Fe sample characterized by the response function $R(\Delta)$. Single photons are then detected by avalanche photodiode detectors (APDs) as a function of time and Doppler detuning. Each analyzer detuning $\Delta_D^k$ probes a local region of $R(\Delta)$, and the sequence of overlapping measurements across multiple detunings provides the redundancy needed to reconstruct both amplitude and phase, in analogy to ptychographic imaging.
		}
		\label{Fig.1} 
	\end{figure*}
	
	Here, we extend ptychographic principles to nuclear spectroscopy and introduce \textbf{Nuclear Ptychoscopy}, a framework that adapts algorithms from coherent diffractive imaging to reconstruct the full complex nuclear response. 
	Our approach establishes a general methodology for diverse experimental conditions, incorporating three complementary reconstruction schemes covering scenarios with known analyzer response, blind reconstruction, and partial prior information. This methodology is realized through three main categories of methods: geometry-based methods, feasible methods, and constrained optimization methods--encompassing tens of distinct ptychographic algorithms and their variants. We further develop geometric analysis tools—including loss landscape visualization, trajectory principal component analysis, and Hessian spectral analysis—to clarify algorithmic behavior. Our framework also integrates data-driven modules, such as the Plug-and-Play (PnP) approach \cite{venkatakrishnan2013plug}, enabling adaptive incorporation of pre-trained denoisers as implicit regularizers. Together, these capabilities offer a clear technical advantage, effectively addressing challenges such as non-convexity, variable coupling, and non-smooth priors. Validated through experimental data and simulations, our framework achieves improved performance compared to recent nuclear phase retrieval methods (e.g., nuclear phase retrieval spectroscopy (NPRS)), including significantly reduced reconstruction errors and enhanced robustness across diverse experimental conditions. These advantages open new possibilities for precision nuclear spectroscopy and metrology. In particular, the enhanced reconstruction accuracy and robustness of Nuclear Ptychoscopy enable more precise determination of nuclear transition linewidths using a broadband input laser, thereby overcoming limitations imposed by the spectral bandwidth of the probing laser. Such capability is crucial for advancing measurements of ultra-narrow nuclear clock transitions~\cite{Shvydko2023,PhysRevLett.132.182501,PhysRevLett.133.013201,Zhang2024,zhang2024229thf4,PhysRevLett.131.263002,cai2021populating,si2025prediction}, whose intrinsic linewidths are orders of magnitude smaller than those of available X-ray or vacuum ultraviolet sources. Nuclear Ptychoscopy thus provides a unified framework that links spectroscopy and ptychography, enabling advanced approaches to precision metrology, coherent control, and quantum technologies in the nuclear regime.
		
		
		
		\section{Mathematical model}
		
		We begin with the experimental setup, shown in Fig.~\ref{Fig.1}. The target contains $^{57}$Fe M\"{o}ssbauer nuclei (14.4 keV resonance, $4.7$ neV linewidth) with response function $R(\Delta)$, where $\Delta$ is the detuning from the nuclear resonance, and is irradiated by a broadband synchrotron X-ray pulse. After passing through the monochromators, the bandwidth of the incident pulse is reduced to the meV range—still orders of magnitude broader than the nuclear transition~\cite{yoda2019x}. A reference analyzer with $^{57}$Fe is mounted on a M\"{o}ssbauer drive, with transmission $T(\Delta-\Delta_D)$, where $\Delta_D$ arises from Doppler detuning. Single photons are detected by an APD array~\cite{baron2006silicon}, which records time-binned counts, while the analyzer’s Doppler detuning, $\Delta_D$, is measured concurrently. The resulting 2D dataset, $I(t,\Delta_D)$, serves as input for NPRS algorithms \cite{Yuan2025}, enabling reconstruction of the full complex, energy-dependent response. Collecting measurements across multiple analyzer detunings naturally provides overlapping information, analogous to ptychography.
		

		
		The measurement process illustrated in Fig.~\ref{Fig.1} can be expressed mathematically as
		\begin{equation}\label{t1}
			I(t,\Delta_D)=\left|\frac{1}{\sqrt{2\pi}} \int_{-\infty}^{\infty}  R(\Delta) T(\Delta-\Delta_D)e^{-i\Delta t}d\Delta\right|^2.
		\end{equation}
		For numerical implementation, Eq.~\eqref{t1} can be discretized as
		\begin{eqnarray}
			I(t_k,\Delta_D^l)=\left|\mathbf{F}_{:,k}^{\text{H}}(\mathbf{R}\odot\mathbf{T}_{l})\right|^{2}+\varepsilon, t_k\in\mathbf{E},\Delta_D^l\in\bm{\Omega},\label{discrete model}
		\end{eqnarray}
		where $\mathbf{F}_{:,k}=\frac{\Delta_{\textrm{step}}}{\sqrt{2\pi}}(1,e^{i1\Delta_{\textrm{step}}\cdot t_k},\cdots,e^{i(n-1)\Delta_{\textrm{step}}\cdot t_k}),$ $\odot$ is the Hadamard product, $(\cdot)^{\text{H}}$ represents the conjugate transpose, and $\varepsilon$ is the error caused by noise and discretization. $\mathbf{R}=(R(\Delta_0),R(\Delta_1),\cdots,R(\Delta_{n-1}))$, and $\mathbf{T}_l=(T(\Delta_0-\Delta_D^l),T(\Delta_1-\Delta_D^l),\cdots,T(\Delta_{n-1}-\Delta_D^l))$, 
		are the discretizations of $R(\Delta)$ and $T(\Delta-\Delta_D)$ in the range $[-\Delta_{\textrm{max}},\Delta_{\textrm{max}}]$ by equally spaced nodes $\Delta_j$ with stepsize $\Delta_{\text{step}}$ under Doppler detunings $\bm{\Omega}=\{\Delta_D^1,\Delta_D^2,\cdots,\Delta_D^L\}$ ordered increasingly, and $\mathbf{E}=\{t_1,t_2,\cdots,t_K\}$ are the sampling points of the time. For each $\mathbf{T}_l$, it can be equivalently formulated as the product of a common $\mathbf{T}\in\mathbb{C}^{n+h}$ and sampling matrix $\mathbf{C}_l\in\mathbb{R}^{n\times(n+h)}$. Specifically, $\mathbf{T}\in\mathbb{C}^{n+h}$ is the extended analyzer response matrix (with $h$ additional nodes to cover the full Doppler detuning range), and $\mathbf{C}_l\in\mathbb{R}^{n\times(n+h)}$ is the sampling matrix that extracts the $n$-dimensional subset of $\mathbf{T}$ corresponding to detuning $\Delta_D^l$. Then, Eq.~\eqref{discrete model} can be transformed as 
		\begin{eqnarray}
			I(k,l)=\left|\mathbf{F}_{:,k}^{\text{H}}\big(\mathbf{R}\odot(\mathbf{C}_{l}\mathbf{T})\big)\right|^{2}+\varepsilon, t_k\in\mathbf{E},\Delta_D^l\in\bm{\Omega},\label{rediscrete model}
		\end{eqnarray}
		where $I(k,l)$ denotes $I(t_k,\Delta_D^l)$ for notational simplicity. From Eq.~\eqref{rediscrete model}, it is evident that for each detuning $\Delta_D^l$, a specific portion of $\mathbf{T}$ interacts with $\mathbf{R}$, producing the time-domain signal $I(t,\Delta_D^l)$ denoted by $I(k,l)$, as illustrated in Fig.~\ref{Fig.1}. This pattern of overlapping measurements is analogous to conventional ptychography \cite{Rodenburg2008}, where a localized, coherent probe scans across a target, generating a sequence of overlapping observations.
		
		The key similarity lies in the presence of redundant, overlapping information, which enables the complex-valued sample response to be reconstructed from intensity-only measurements. The main difference is that, in our experiment, the sample $\mathbf{R}$ remains fixed, while the Doppler-tuned analyzer $\mathbf{T}$ acts as the “probe.” In this sense, our setup implements a conjugate form of ptychography, effectively swapping the roles of probe and object compared to traditional spatial scanning.

		By performing a change of variables in Eq.~\eqref{t1}, the measurement can be rewritten in a form directly analogous to ptychography:
		\begin{eqnarray}\label{t2}
			I(t,\Delta_D)=\left|\frac{1}{\sqrt{2\pi}} \int_{-\infty}^{\infty}  R(\Delta+\Delta_D) T(\Delta)e^{-i\Delta t}d\Delta\right|^2.
		\end{eqnarray}
		Similarly, its discrete representation is
		\begin{eqnarray*}
			I(t_k,\Delta_D^l)=\left|\mathbf{F}_{:,k}^{\text{H}}\big((\mathbf{C}_l\mathbf{R})\odot\mathbf{T}\big)\right|^{2}+\varepsilon, t_k\in\mathbf{E},\Delta_D^l\in\bm{\Omega},\label{discrete model1}
		\end{eqnarray*}
		where $\mathbf{R}\in\mathbb{C}^{n+h}$, and $\mathbf{T}\in\mathbb{C}^n$. This formulation casts nuclear resonant scattering with a Doppler-tuned analyzer into a ptychographic framework, providing the conceptual basis of our Nuclear Ptychoscopy approach, which enables simultaneous reconstruction of both the amplitude and phase of the nuclear response.
		\section{The algorithms of Nuclear Ptychoscopy}
		With the concept of Nuclear Ptychoscopy established, we now turn to its practical reconstruction workflows. We present three complementary reconstruction schemes tailored to distinct experimental constraints: (i) reconstruction with a known analyzer response, suitable for well-calibrated setups, (ii) blind reconstruction, for scenarios where the analyzer response cannot be characterized, and (iii) reconstruction incorporating partial prior information, which leverages limited knowledge of the analyzer or target to reduce ambiguity and improve reconstruction accuracy. These schemes are realized through ten distinct ptychographic algorithms, summarized in Table~\ref{TB}.
		
		Building on the analysis of our model and its relation to conventional ptychography, we next evaluate the ten distinct ptychographic algorithms summarized in Table~\ref{TB} to assess their suitability for each of the three reconstruction schemes in Nuclear Ptychoscopy.
		\begin{table*}
			\caption{Summary of the algorithms in this work, systematically classified into three distinct categories. Details of the variables and notations are available in the corresponding sections.}
			\label{TB}
			\centering
			\begin{tabular}{|l|l|l|}
				\hline 
				\multicolumn{1}{|c|}{Categories}& \multicolumn{1}{c|}{Names of Algorithm} & \multicolumn{1}{c|}{Core iterative procedures in iteration $m$}  \\ 
				\hline
				\multicolumn{1}{|c|}{\multirow{4}{*}{\begin{tabular}[c]{@{}c@{}}\\\\\\~\\\\\\\\Geometry-based \\ Methods\end{tabular}}}
				&\multicolumn{1}{c|}{\tabincell{c}{NPRS(Nuclear Phase\\ Retrieval Spectroscopy)}}&
				\tabincell{l}{\\
					1. $\tilde{\mathbf{R}}^{(m)} = \mathbf{R}^{(m)} -\lambda^{(m)}\nabla \ell(\mathbf{R}^{(m)})$\\
					2. $\mathbf{R}^{(m + 1)} = \tilde{\mathbf{R}}^{(m)}+\frac{t-1}{t+2}(\tilde{\mathbf{R}}^{(m)}-\tilde{\mathbf{R}}^{(m-1)})$\\
					3. If $\ell(\mathbf{R}^{(m+1)})<\ell(\mathbf{R}^{(m)})$:~~$t=t+1$ else: $t=1$\\
					\\
				}
				\\ \cline{2-3} 
				&\multicolumn{1}{c|}{\tabincell{c}{Nonlinear Conjugate\\ Gradient(NCG)}}& 
				\tabincell{l}{
					\\
					1. $\mathbf{D}^{(m)}=-\nabla \ell(\mathbf{R}^{(m)})+\left(\frac{\operatorname{Re}\left[(\nabla\ell(\mathbf{R}^{(m)}))^{\textrm{H}}\left(\nabla\ell(\mathbf{R}^{(m)})-\nabla\ell(\mathbf{R}^{(m-1)})\right)\right]}{\left\|\nabla\ell(\mathbf{R}^{(m-1)})\right\|^2_2}\right) \mathbf{D}^{(m-1)}$\\
					2. $\mathbf{R}^{(m + 1)}=\mathbf{R}^{(m)}+\lambda^{(m)}\mathbf{D}^{(m)}$\\
					\\
				}
				\\ \cline{2-3} 
				& \multicolumn{1}{c|}{Levenberg-Marquardt(LM)}&
				\tabincell{l}{
					\\
					1. $J(\mathbf{R}^{(m)})=\nabla r(\mathbf{R}^{(m)})^{\text{H}}\nabla r(\mathbf{R}^{(m)})$\\
					~~~~where $r(\mathbf{R}) = \frac{1}{\sqrt{2}}\left(|\bm{\Psi}|^p-\mathbf{I}^{\frac{p}{2}}\right)$,~$\Psi(k,l) = \mathbf{F}^{\text{H}}_{:,k}(\mathbf{R}\odot\mathbf{T}_l)$,~$p=1$ or $2$\\
					2. $\mathbf{D}^{(m)}\in\arg\min\limits_{\mathbf{D}\in\mathbb{C}^n}\|\left(J(\mathbf{R}^{(m)})+\rho^{(m)}\bm{1}_{n\times n}\right)\mathbf{D}+\nabla\ell(\mathbf{R}^{(m)})\|^2_2$\\
					3. $\mathbf{R}^{(m+1)}=\mathbf{R}^{(m)}+\mathbf{D}^{(m)}$\\
					\\
				}
				\\ \cline{2-3} 
				& \multicolumn{1}{c|}{\tabincell{c}{Limited-Memory Broyden-\\Fletcher-Goldfarb-Shanno\\(L-BFGS)}}&
				\tabincell{l}{\\
					
					1. If $|\mathcal{P}|>q$: $\mathcal{P}=(\mathcal{P}\setminus\{\mathbf{R}^{(m-q)}\})\cup\{\mathbf{R}^{(m)}\}$,\\
					~~~~~~~~~~~~~~and $\mathcal{G}=\left(\mathcal{G}\setminus\{\nabla \ell(\mathbf{R}^{(m-q)})\}\right)\cup\{\nabla \ell(\mathbf{R}^{(m)})\}$ 
					\\  
					~~~~else: $\mathcal{P}=\mathcal{P}\cup\{\mathbf{R}^{(m)}\}$, $\mathcal{G}=\mathcal{G}\cup\{\nabla\ell(\mathbf{R}^{(m)})\}$\\
					2. $\mathbf{D}^{(m)}=\textit{ L-BFGS two loop recursion}\left(\nabla \ell(\mathbf{R}^{(m)}),\mathcal{P},\mathcal{G}\right)$\\
					3. $\mathbf{R}^{(m + 1)}=\mathbf{R}^{(m)}+\lambda^{(m)}\mathbf{D}^{(m)}$\\
					\\
				}
				\\ 
				\hline
				\multicolumn{1}{|c|}{\multirow{3}{*}{{\tabincell{c}{\\\\\\\\Feasible Methods}}}
				}&\multicolumn{1}{c|}{\tabincell{c}{AP(Alternating Projection)}}&
				\tabincell{l}{\\
					1. $\bm{\Psi}^{(m+1)}=\mathbb{P}_{\mathcal{B}}\left(\mathbb{P}_{\mathcal{A}}(\mathbf{\Psi}^{(m)})\right)$,~$\Psi^{(m)}(k,l) = \mathbf{F}^{\text{H}}_{:,k}(\mathbf{R}^{(m)}\odot\mathbf{T}_l)$\\\\
					2.~If $\{\mathbf{T}_l\}_{1\leq l\leq L}$ are known: $\mathbf{R}^{(m+1)}=\frac{\sum_{l=1}^L \overline{\mathbf{T}_l} \odot \mathbf{\Psi}_l^{(m+1)}}{\sum_{l=1}^L\left|\mathbf{T}_l\right|^2}$
					\\
					~~~~else: $\mathbf{R}^{(m+1)}=\frac{\sum_{l=1}^L \overline{\mathbf{C}_l\mathbf{T}^{(m)}} \odot \mathbf{\Psi}_l^{(m+1)}}{\sum_{l=1}^L\left|\mathbf{C}_l\mathbf{T}^{(m)}\right|^2}$ \\
					~~~~~~~~~~~$\mathbf{T}^{(m+1)}\in\arg \min \limits_{\mathbf{T} \in \mathbb{C}^{n+h}} \frac{1}{2} \sum_{l=1}^L\left\|\bm{\Psi}_l^{(m+1)}-\mathbf{R}^{(m+1)} \odot\left(\mathbf{C}_l \mathbf{T}\right)\right\|^2_2$\\
					\\
				}
				\\ 
				\cline{2-3} 
				&\multicolumn{1}{c|}{\tabincell{c}{DR(Douglas Rachford)}}& 
				\tabincell{l}{\\
					1. Same with AP, except $\bm{\Psi}^{(m+1)}=\frac{1}{2}\left(\mathbb{R}_{\mathcal{B}}\mathbb{R}_{\mathcal{A}}+\mathbb{I}\right)(\mathbf{\Psi}^{(m)})$\\
					\\
				}
				\\ \cline{2-3} 
				&\multicolumn{1}{c|}{\tabincell{c}{RAAR(Relaxed Averaged\\ Alternating Reflections )}}&
				\tabincell{l}{\\
					1. Same with AP, except\\ ~~ $\bm{\Psi}^{(m+1)}=\frac{\alpha}{2}\left(\mathbb{R}_{\mathcal{B}}\mathbb{R}_{\mathcal{A}}+\mathbb{I}\right)(\mathbf{\Psi}^{(m)})+(1-\alpha)\mathbb{P}_{\mathcal{A}}(\mathbf{\Psi}^{(m)})$, $0\leq\alpha\leq1$\\
					\\
				}
				\\
				
				\hline
				\multirow{3}{*}{\tabincell{c}{\\\\\\\\\\\\Constrained Optimi-\\zation Methods}} &\multicolumn{1}{c|}{\tabincell{c}{Proximal gradient method}}   & 
				\tabincell{l}{\\
					1. $\tilde{\mathbf{R}}^{(m)} \in \arg\min\limits_{\mathbf{R}\in\mathbb{C}^n}\frac{1}{2\lambda}\|\mathbf{R}-\mathbf{R}^{(m)} +\lambda\nabla \ell(\mathbf{R}^{(m)})\|^2_2+\tau\mathcal{R}(\mathbf{R})$\\
					2. $t^{(m)}=\frac{1}{2}\left(1+\sqrt{4(t^{(m-1)})^2-1}\right)$\\
					3. $\mathbf{R}^{(m+1)}=\tilde{\mathbf{R}}^{(m)}+\frac{t^{(m-1)}-1}{t^{(m)}}(\tilde{\mathbf{R}}^{(m)}-\tilde{\mathbf{R}}^{(m-1)})$\\
					\\
				}
				\\ \cline{2-3} 
				&\multicolumn{1}{c|}{\tabincell{c}{ADMM(Alternating\\ Direction method\\ of Multiplier)}}&
				\tabincell{l}{\\
					1. $\mathbf{R}^{(m+1)}\in\arg\min\limits_{\mathbf{R}\in\mathbb{C}^n}\ell(\mathbf{R})+\frac{\eta}{2}\|\mathbf{R}-\mathbf{Y}^{(m)}+\mathbf{Z}^{(m)}\|^2_2$\\
					2. $\mathbf{Y}^{(m+1)}\in\arg\min\limits_{\mathbf{Y}\in\mathbb{C}^n}\frac{\eta}{2}\|\mathbf{Y}-\mathbf{R}^{(m+1)}-\mathbf{Z}^{(m)}\|^2_2+\tau\mathcal{R}(\mathbf{Y})$\\
					3. $\mathbf{Z}^{(m+1)} = \mathbf{Z}^{(m)}+\lambda(\mathbf{R}^{(m+1)}-\mathbf{Y}^{(m+1)})$\\
					
					\\
				}
				\\
				\cline{2-3} 
				&\multicolumn{1}{c|}{\tabincell{c}{Constrained RAAR}}&
				\tabincell{l}{\\
					1. Same with RAAR, except\\ 
					~~ $ \mathbf{R}^{(m+1)}\in\arg \min \limits_{\mathbf{R}\in\mathbb{C}^n}\sum_{l=1}^L\left\|\mathbf{\Psi}_l^{(m+1)}-\mathbf{R} \odot\left(\mathbf{C}_l \mathbf{T}^{(m)}\right)\right\|^2_2+\tau_1\mathcal{R}_1(\mathbf{R})$ \\
					~~ $\mathbf{T}^{(m+1)}\in\arg \min\limits_{\mathbf{T}\in\mathbb{C}^{n+h}} \sum_{l=1}^L\left\|\mathbf{\Psi}_l^{(m+1)}-\mathbf{R}^{(m+1)} \odot\left(\mathbf{C}_l \mathbf{T}\right)\right\|^2_2+\tau_2\mathcal{R}_2(\mathbf{T})$\\
					\\
				}\\
				\hline
			\end{tabular}
		\end{table*}
		
		\subsection{Geometry-based methods}
		
		As the first category within Nuclear Ptychoscopy, geometry-based methods address the scenario where the analyzer response is fully known. In this case, the reconstruction problem can be formulated as
		\begin{equation}
			\begin{gathered}
				\text{Find}~\mathbf{R}\in\mathbb{C}^n\\
				I(k, l)=\left|\mathbf{F}_{:, k}^{\mathrm{H}}\left(\mathbf{R} \odot \mathbf{T}_l\right)\right|^2+\varepsilon, \\
				k \in\{1,2, \cdots, K\}, l \in\{1,2, \cdots, L\},
			\end{gathered}\label{nps}
		\end{equation}
		which seeks to recover the complex-valued nuclear response $\mathbf{R}$ from intensity-only measurements. Addressing this problem requires the definition of a suitable loss function and its gradient, followed by the development of an algorithm to find its minimizer. In the following, we present three distinct loss–gradient formulations, each giving rise to a corresponding reconstruction algorithm tailored to this experimental scenario.
		
		\subsubsection{The loss function}
		Following the Bayesian formulation in Ref.~\cite{Yuan2025}, the reconstruction of the nuclear response $\mathbf{R}$ from measured intensities $\mathbf{I}$ can be expressed as
		\begin{eqnarray}\label{bayesian}
			\mathbf{P}(\mathbf{R}|\mathbf{I})\propto\mathbf{P}(\mathbf{I}|\mathbf{R})\mathbf{P}(\mathbf{R}).
		\end{eqnarray}
		In the general case, the prior term $\mathbf{P}(\mathbf{R})$ is usually omitted, and the reconstruction is based on maximizing the likelihood function $\mathbf{P}(\mathbf{I}|\mathbf{R})$.
		In optical experiments, the Poisson model is often appropriate, yielding a loss function based on Poisson maximum-likelihood estimation. From a mathematical viewpoint, however, alternative statistical models can also be justified. For instance, a Gaussian model formulated in terms of amplitude or intensity leads to distinct least-squares estimators. Thus, the optimization may be performed using any of the following three loss functions,
		\begin{equation*}
			\ell(\mathbf{R})=\left\{    
			\begin{aligned}
				&\underbrace{\frac{1}{KL}\sum\limits_{k=1}^{K}\sum\limits_{l=1}^{L}\left(|\Psi(k,l)|^2-I(k,l)\log|\Psi(k,l)|^2\right)}_{\text{(Poisson Model)}}\\&\underbrace{\frac{1}{2KL}\sum\limits_{k=1}^{K}\sum\limits_{l=1}^{L}\left(|\Psi(k,l)|^2-I(k,l)\right)^2}_{\text{(Intensity-based Gaussian Model)}}\\
				&\underbrace{\frac{1}{2KL}\sum\limits_{k=1}^{K}\sum\limits_{l=1}^{L}\left(|\Psi(k,l)|-\sqrt{I(k,l)}\right)^2}_{\text{(Amplitude-based Gaussian Model)}}
			\end{aligned}\right.,
		\end{equation*}
		where $\Psi(k,l) = \mathbf{F}^{\text{H}}_{:,k}(\mathbf{R}\odot\mathbf{T}_l)$. It should be noted that the Poisson and amplitude-based Gaussian models yield a quadratic formulation of the phase-retrieval problem, whereas the intensity-based Gaussian model leads to a quartic formulation. As shown in Ref.~\cite{chen2017solving, wang2017solving}, algorithms applied to quadratic formulations generally achieve superior performance in terms of both reconstruction accuracy and convergence reliability. Consistent with these findings, we observe a similar trend in Nuclear Ptychoscopy: as illustrated in Fig.~\ref{Fig.2}, algorithms such as NPRS, nonlinear conjugate gradient (NCG), Levenberg–Marquardt (LM), and limited-memory Broyden–Fletcher–Goldfarb–Shanno (L-BFGS) attain better performance with the Poisson and amplitude-based Gaussian models than with the intensity-based Gaussian model. Details will be presented in the following sections.
		
		Given the prohibitive computational expense associated with the direct calculation of the Hessian matrix and its inverse, the gradient information of the loss function is often preferred. Accordingly, the Wirtinger derivatives $\nabla\ell(\mathbf{R})$ (we simplify $\nabla_{\mathbf{R}}(\cdot)$ as $\nabla(\cdot)$) for the respective models are derived as follows via application of the chain rule:
		\begin{equation*}
			\left\{    
			\begin{aligned}
				&\underbrace{\sum\limits_{k=1}^{K}\sum\limits_{l=1}^{L}\left(\Psi(k,l)-\frac{I(k,l)\Psi(k,l)}{|\Psi(k,l)|^2}\right)\frac{\nabla\Psi(k,l)}{KL}}_{\text{(Poisson Model)}}\\&\underbrace{\sum\limits_{k=1}^{K}\sum\limits_{l=1}^{L}\left(|\Psi(k,l)|^2-I(k,l)\right)\frac{\nabla\Psi(k,l)\Psi(k,l)}{KL}}_{\text{(Intensity-Based Gaussian Model)}}\\
				&\underbrace{\sum\limits_{k=1}^{K}\sum\limits_{l=1}^{L}\left(\Psi(k,l)-\frac{\sqrt{I(k,l)}\Psi(k,l)}{|\Psi(k,l)|}\right)\frac{\nabla\Psi(k,l)}{KL}}_{\text{(Amplitude-based Gaussian Model)}}
			\end{aligned}\right.,
		\end{equation*}
		where $\nabla\Psi(k,l)=\mathbf{F}_{:,k}\odot\overline{\mathbf{T}}_l$, and $\overline{\cdot}$ is the conjugate operator by element.
		\subsubsection{Algorithms}
		As demonstrated in Ref.~\cite{Yuan2025}, NPRS method takes advantage of the gradient derived from a Poisson likelihood model, enhanced with acceleration techniques, weighting schemes, and restart strategies, to converge toward the ground truth solution. The NPRS method can recover both the intensity and the phase of the sample with high accuracy in simulations and experiments. Beyond first-order gradient information, the convergence properties of optimization methods can be significantly enhanced by leveraging conjugate gradient directions and incorporating second-order information from the Hessian matrix. To this end, here we apply three advanced optimization techniques—the NCG method, the LM method, and the L-BFGS method to solve the inverse problem posed by the Nuclear Ptychoscopy model. The main steps for each of these algorithms are summarized in Table \ref{TB}.
		
		To achieve a better performance, all optimization methods are tailored to the specific characteristics of Nuclear Ptychoscopy. For both the NCG, LM and L-BFGS methods, the loss function is selected as either the amplitude(intensity)-based Gaussian model or the Poisson likelihood model, constrained by its underlying formulation. Consequently, Poisson model is not suitable for LM method. And the step size $\lambda^{(m)}$ are estimated by using a backtracking line search strategy. In the case of NCG, the Polak–Ribière–Polyak (PRP) conjugate direction is employed to update the search direction at each iteration.
		
		\begin{figure*}[ht]
			\setlength{\abovecaptionskip}{-0.2cm} 
			\centerline{\includegraphics[width=1.00\linewidth]{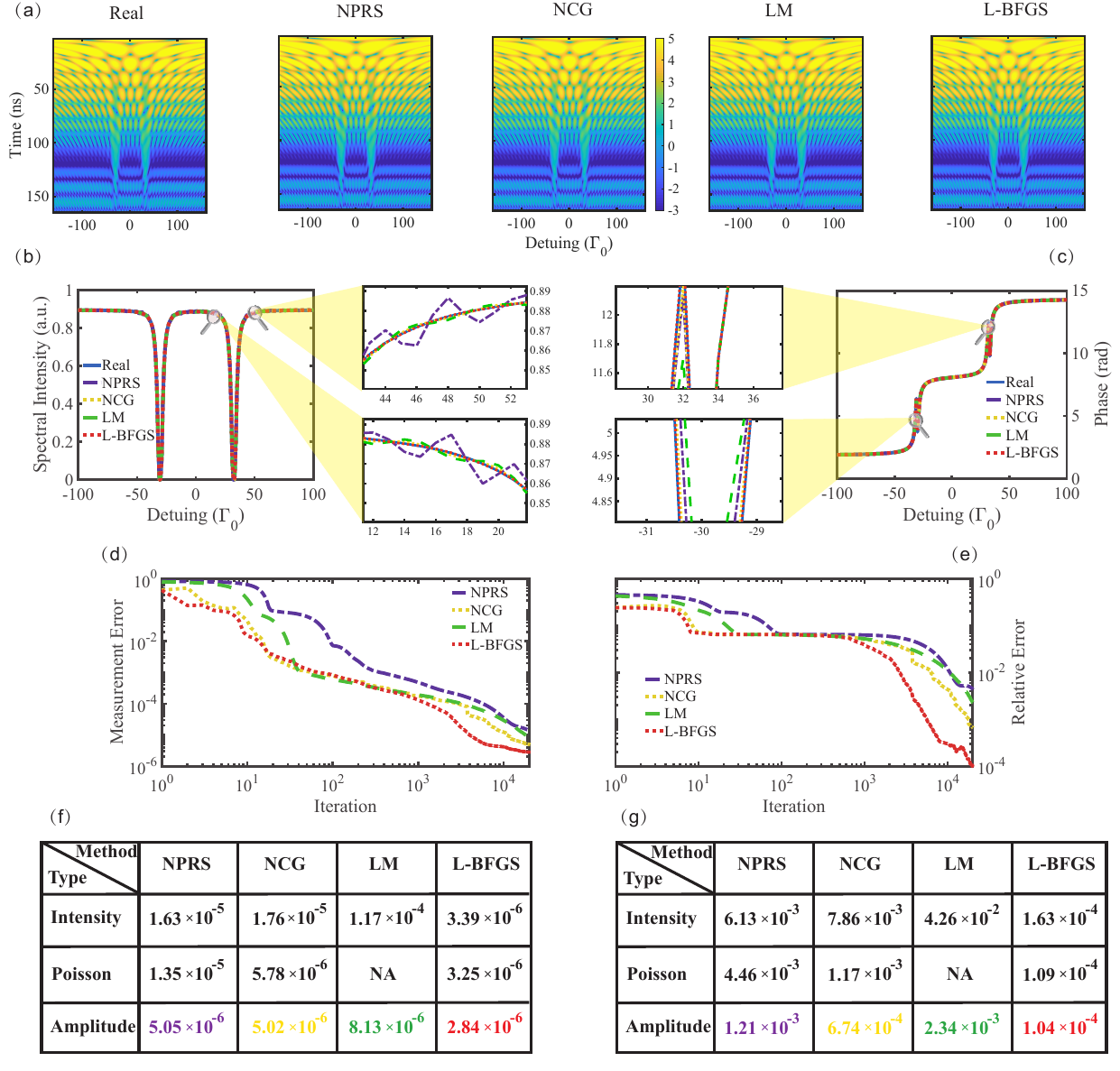}}
			\centering
			\begin{picture}(10,5)
			\end{picture}
			\caption{Comparative performance of optimization algorithms in Nuclear Ptychoscopy reconstruction. (a) 2D  spectrum: ground truth (labeled "Real") and reconstructions from the NPRS, NCG, LM, and L-BFGS methods. (b) Recovered intensity and (c) its corresponding phase. Insets in (b) and (c) show magnified views highlighting key structural details. (d) Measurement error (defined in Eq. \eqref{mtt21}) and (e) relative error (defined in Eq. \eqref{mtt22}), plotted against iteration count for each algorithm. (f) and (g) respectively record the measurement error and relative error of the four algorithms under three different loss functions: the Poisson model, the intensity-based Gaussian model, and the amplitude-based Gaussian model. }
			\label{Fig.2} 
		\end{figure*}
		To comprehensively evaluate the performance of the NPRS, NCG, LM, and L-BFGS algorithms, we consider the numerical case presented in Ref.~\cite{Yuan2025}, where a linearly polarized X-ray pulse irradiates an $\alpha$-$^{57}\text{Fe}$ target with an effective thickness of $d=2.3\,\mu$m under normal incidence. A single-resonance $\mathrm{K}_2\mathrm{Mg}^{57}\mathrm{Fe}(\mathrm{CN})_6$ analyzer with an effective thickness of $d\approx1\,\mu$m is included, as in the experiment. To reproduce the experimental conditions at the SR beamline, the time window is set from 3~ns to 165~ns. Signals before 3~ns are excluded due to the strong prompt off-resonant component of the incident X-ray pulse, while the upper limit of 165~ns corresponds to the bunch separation of the E mode at the nuclear resonant scattering beamline BL35XU of SPring-8, Japan. The numerical results are shown in Fig.~\ref{Fig.2}.
		
		The results presented in Fig.~\ref{Fig.2}(f) and (g) quantitatively compare the performance of four optimization algorithms—NPRS, NCG, LM, and L-BFGS—across three distinct loss functions: the intensity-based Gaussian model, the Poisson model, and the amplitude-based Gaussian model. To quantify these differences, we introduce two error metrics. The measurement error is defined as
		\begin{eqnarray}
			\sqrt{\frac{\sum_{k=1}^{K}\sum_{l=1}^{L}	\left(I(k,l)-\left|\mathbf{F}_{:,k}^{\text{H}}(\tilde{\mathbf{R}}\odot\mathbf{T}_{l})\right|^{2}\right)^2}{\sum_{k=1}^{K}\sum_{l=1}^{L}I^2(k, l)}},
			\label{mtt21}
		\end{eqnarray}
		characterizing deviations in the 2D spectrum. The relative error is defined as
		\begin{eqnarray}
			\sqrt{\frac{\sum_{\Delta\in\mathbf{O}}	\left|\tilde{\mathbf{R}}(\Delta)e^{i\theta}-\mathbf{R}(\Delta)\right|^2}{\sum_{\Delta\in\mathbf{O}}\left|\mathbf{R}(\Delta)\right|^2}},
			\label{mtt22}
		\end{eqnarray}
		quantifying deviations in the reconstructed nuclear response spectrum, where $\tilde{\mathbf{R}}$ denotes the reconstructed nuclear response, and $\theta$ is a global phase correction term (accounting for the inherent global phase ambiguity in phase retrieval \cite{Yuan2025}).
		
		We find that the amplitude-based Gaussian model consistently yields the lowest errors among all methods. This observation aligns with the findings presented in Ref.~\cite{yeh2015experimental}, which highlights the advantages of amplitude-based Gaussian formulations due to their more favorable geometric structure and better-behaved gradients in Fourier ptychography. Although the Poisson model also demonstrates competitive performance, it is not applicable to the LM method in this particular configuration. Subsequent analysis focuses on a comparison of the four algorithms under the \textbf{amplitude-based Gaussian model}. It should be noted that while the NPRS method was originally introduced in Ref.~\cite{Yuan2025} for solving the Poisson model, we adapt it here to the amplitude-based Gaussian formulation by modifying its gradient calculation to match the amplitude-based loss function for a consistent comparative framework.
		
		Visually, as shown in Fig.~\ref{Fig.2}(a)–(c), the reconstructed 2D spectrum, together with the intensity and phase of the nuclear response function, agree closely with the ground truth across all methods. In this work, the 2D spectrum labeled “Real” denotes the ground-truth spectrum used in simulations, while the one labeled “Theory” represents the theoretically calculated spectrum obtained using parameters extracted from fitting the independently measured time spectrum. The 2D spectrum labeled “Measured” corresponds to the experimental data acquired by the APDs. Meanwhile, the same labeling conventions are applied to their corresponding intensities and phases of the target spectrum. Detailed observations of zoomed-in regions, however, reveals differences in reconstruction quality. The L-BFGS method yields the most accurate result, followed by the NCG method. Compared to the NPRS method, the LM algorithm achieves a better fit to the intensity spectrum but performs less accurately in phase recovery.
		
		Measurement and relative errors, tracked throughout the optimization, are presented in Fig.~\ref{Fig.2}(d) and (e). Consistent with the qualitative observations, the L-BFGS method demonstrates superior performance, exhibiting the fastest convergence rate along with the lowest final measurement and relative errors. The NCG method ranks second in performance. 
		
		Although the phase reconstructed by the LM method does not exhibit the same visual fidelity to the ground truth as other methods, it achieves lower relative and measurement errors than the NPRS approach. This result indicates that the LM method demonstrates superior performance compared to NPRS in terms of quantitative accuracy.
		
\begin{figure*}[ht]
			\setlength{\abovecaptionskip}{-0.2cm} 
			\centerline{\includegraphics[width=1.00\linewidth]{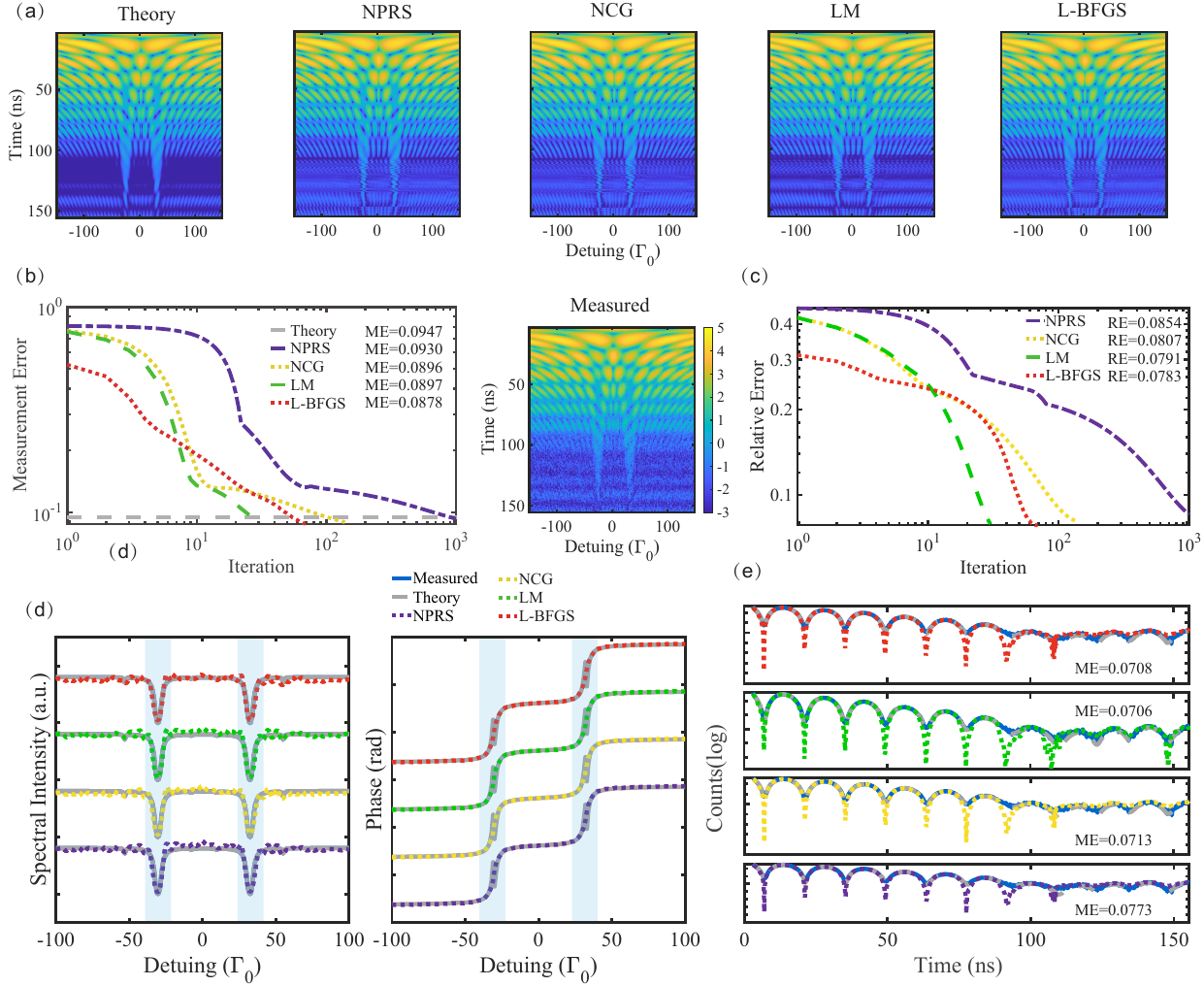}}
			\centering
			\begin{picture}(10,5)
			\end{picture}
			\caption{Comparative performance of geometry-based optimization methods on experimental data. (a) Measured 2D  spectrum alongside its theoretical prediction and reconstructions from the NPRS, NCG, LM, and L-BFGS algorithms. (b) Measurement error and (c) relative error, as defined in Eqs.\eqref{mtt21} and \eqref{mtt22}, plotted as functions of iteration number. (d) Reconstructed intensity and phase of the target. (e) Time spectra obtained from Fourier transforms of the reconstructed amplitude and phase for each method, compared with independently measured experimental data, demonstrating reconstruction accuracy. The corresponding measurement error is shown in each plot. Here, the spectrum (2D spectrum, intensity or phase spectrum) labeled “Theory” represents the theoretically calculated spectrum obtained using parameters extracted from fitting the independently measured time spectrum. The 2D spectrum labeled “Measured” corresponds to the experimental data acquired by the APDs.}
			\label{Fig.3} 
		\end{figure*}

		To empirically validate the performance of the aforementioned optimization methods—NPRS, NCG, LM, and L-BFGS, we utilized experimental data presented in Ref.~\cite{Yuan2025}, which is measured at the nuclear resonant scattering beamline BL35XU of SPring-8 facility in Japan. As shown in Fig.~\ref{Fig.3}(a), the results demonstrate that all four methods successfully reconstruct the two-dimensional measurements, with outputs aligning closely with theoretical predictions and experimental observations. 
		
		
		Furthermore, the recovered intensity spectrum and corresponding phase consistently conform to theoretical expectations, as shown in  Fig.~\ref{Fig.3}(d), reinforcing the applicability of geometry-based computational methods for solving the Nuclear Ptychoscopy problem under the condition of a known analyzer $\mathbf{T}$. A quantitative analysis based on measurement error and relative error metrics—defined in Eqs.~\eqref{mtt21} and \eqref{mtt22}—reveals that the L-BFGS method achieves the lowest errors, showing its superior performance in terms of both reconstruction accuracy and convergence efficiency, as demonstrated in Fig.~\ref{Fig.3}(b) and (c). Notably, the LM method also demonstrates competitive performance, outperforming both NCG and NPRS in terms of the final measurement and relative errors, though it remains secondary to L-BFGS overall. As an additional validation, the time-dependent spectra retrieved by the above methods exhibit excellent agreement with both the measured data points and their fits, as shown in Fig.~\ref{Fig.3}(e). Specifically, the LM method yields the lowest measurement error for the time spectra of the recovered $\mathbf{R}$, while the measurement error from the L-BFGS method is nearly identical to this value, and the NCG method follows next.   
		
		Overall, geometry-based optimization methods exhibit a clear performance hierarchy, with second-order methods, namely NCG, LM, and L-BFGS, demonstrating significant advantages over the first-order NPRS approach. The superior performance of L-BFGS can be attributed to its effective incorporation of approximate Hessian information via quasi-Newton updates (using limited historical gradient data to approximate curvature), enabling faster convergence and higher reconstruction accuracy. Both NCG and LM also deliver competitive results, offering a favorable balance between performance and memory efficiency, as they require less storage for historical gradient or curvature information compared to L-BFGS.
		
		Despite its comparatively lower accuracy, the NPRS method remains notable for its straightforward implementation and computational efficiency, often yielding satisfactory results with minimal tuning. This highlights a trade-off between computational complexity and reconstruction quality, allowing flexibility in choice depending on resource availability and accuracy requirements.
		
		While geometry-based methods demonstrate good performance when the analyzer response $\mathbf{T}$ is known, 
		such prior knowledge is not always available in realistic experimental settings (e.g., uncalibrated analyzers). In such cases, geometry-driven methods fail to simultaneously reconstruct $\mathbf{R}$ and $\mathbf{T}$, making their limitations apparent and necessitating alternative strategies such as blind reconstruction.
		
		\subsection{Feasible methods for blind Nuclear Ptychoscopy}
		
\begin{figure*}[ht]
			\setlength{\abovecaptionskip}{-0.2cm} 
			\centerline{\includegraphics[width=1.00\linewidth]{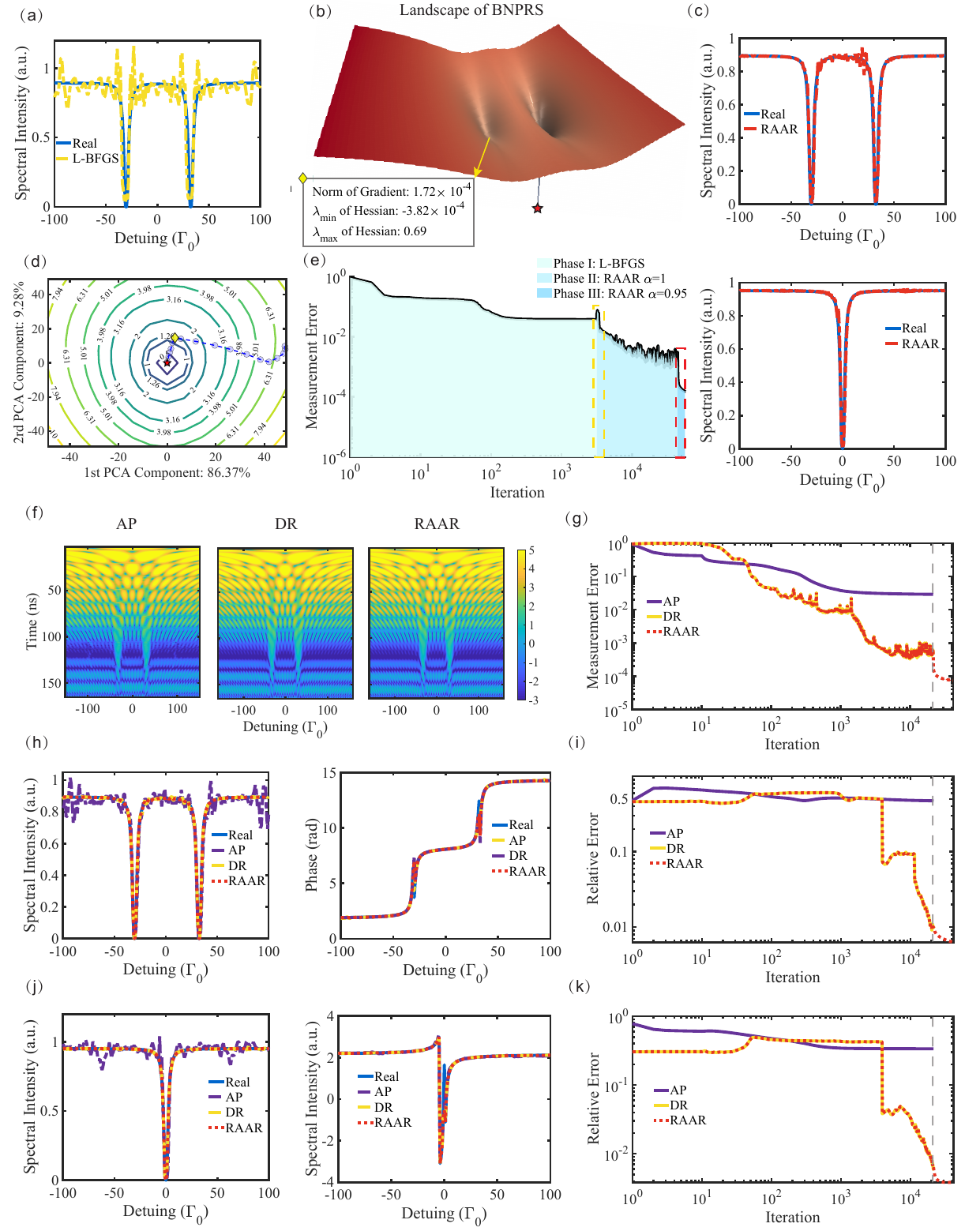}}
			\centering
			\begin{picture}(10,5)
			\end{picture}
			\caption{Comprehensive analysis of feasible methods. (a) Intensity reconstructed via the L-BFGS method. (b) Loss landscape of the optimization problem, which is rendered by the \href{http://paraview.org}{ParaView}. (c) Intensity reconstructed by the RAAR method, initialized at the L-BFGS stagnation point. (d) Iteration trajectories of L-BFGS and RAAR, with corresponding measurement errors in (e), where the curve consists of the results generated in three phases, namely the L-BFGS method, the RAAR method when $\alpha=1$, and the RAAR method when $\alpha=0.95$. (f) Recovered 2D spectrum using AP, DR and RAAR. (g) Mmeasurement error of the 2D spectrum. (h) Recovered intensity and phase of $\mathbf{R}$. (i) Relative error of $\mathbf{R}$ during iteration. (j) Recovered intensity and phase of $\mathbf{T}$. (k)  Relative error of $\mathbf{T}$ during iteration. }
			\label{Fig.4} 
		\end{figure*}

		The performance of geometry-based algorithms, while effective in Nuclear Ptychoscopy, relies critically on prior knowledge of the analyzer response $\mathbf{T}$. When $\mathbf{T}$ is unknown, their efficacy deteriorates markedly, as shown in Fig.~\ref{Fig.4}(a). This situation parallels the blind ptychography problem, where the probe is unknown and has been successfully addressed using feasible methods. Motivated by this analogy, we extend the feasible-method framework to Nuclear Ptychoscopy with an unknown analyzer response. The reconstruction problem in this scenario can be formulated as
		\begin{equation}
			\begin{gathered}
				\text { Find } \mathbf{R} \in \mathbb{C}^n, \mathbf{T}_l \in \mathbb{C}^{n+h},l=1,2,\cdots,L \\
				\text { s.t. } I(k, l)=\left|\mathbf{F}_{:, k}^{\mathrm{H}}\left(\mathbf{R} \odot\mathbf{T}_l\right)\right|^2+\varepsilon, \\
				k=1, \cdots, K, l=1, \cdots, L .\label{BNPY}
			\end{gathered}
		\end{equation}
		
		To establish a mathematical framework for the blind Nuclear Ptychoscopy problem, we derive its formulation from the discrete model in
		Eq.~\eqref{discrete model}. Assume that $\Delta_D^1, \Delta_D^2, \cdots, \Delta_D^L$, which are the multiples of $\Delta_{\text {step }}$, are ordered decreasingly. Let $\Delta_D^{1^{+}}=\max \left(\Delta_D^1, 0\right), \Delta_D^{L^{-}}=\min \left(\Delta_D^L, 0\right)$, and $h:=\frac{\Delta_D^{1^{+}}-\Delta_D^{L^{-}}}{\Delta_{\mathrm{step}}}$, then define a vector $\mathbf{T} \in \mathbb{C}^{n+h}$, which has the formulation below
		$$
		\begin{aligned}
			\mathbf{T} &= \left(T\left(\Delta_0-\Delta_D^{1^{+}}\right),\ T\left(\Delta_0-\Delta_D^{1^{+}}+\Delta_{\text {step }}\right),\right. \\
			&\quad \left.T\left(\Delta_0-\Delta_D^{1^{+}}+2 \Delta_{\text {step }}\right),\ \cdots,\ T\left(\Delta_{n-1}-\Delta_D^{L^{-}}\right)\right).
		\end{aligned}
		$$
		Meanwhile, we also construct a set of matrices $\left\{\mathbf{C}_l \in \mathbb{C}^{n \times(n+h)}, l=1,2, \cdots, L\right\}$ as below (the row and column indices are given in blue for orientation),
		\begin{equation*}
			\setlength{\arraycolsep}{2pt} 
			\renewcommand{\arraystretch}{1} 
			\mathbf{C}_l = 
			\left(
			\begin{array}{c|cccccccccc}
				& \textcolor{blue}{1} & \textcolor{blue}{2} &\textcolor{blue}{\dots} & \textcolor{blue}{f_l+1} & \textcolor{blue}{f_l+2} & \textcolor{blue}{\dots} & \textcolor{blue}{f_l+n} & \textcolor{blue}{\dots} &\textcolor{blue}{ h+n} \\
				\hline
				\textcolor{blue}{1} & 0 & 0 & \dots & 1 & 0 & \dots & 0 & \dots & 0 \\
				\textcolor{blue}{2} & 0 & 0 & \dots & 0 & 1 & \dots & 0 & \dots & 0 \\
				\textcolor{blue}{\vdots} & \vdots & \vdots & \dots & \vdots & \vdots & \ddots & \vdots & \dots & \vdots \\
				\textcolor{blue}{n} & 0 & 0 & \dots & 0 & 0 & \dots & 1 & \dots & 0
			\end{array}
			\right),
		\end{equation*}
		where $f_l=\frac{\Delta_D^{1^+}-\Delta_D^l}{\Delta_{\text {step }}}, l=1,2, \cdots, L$, and the matrix elements in $\left(f_l+i, i\right), i=1,2, \cdots, n$, are 1 , with all other matrix elements 0 . Then, we define $\mathbf{T}_l:=\mathbf{C}_l \mathbf{T}$. With the definition of $\mathbf{C}_l$, the analyzer-free model Eq.~\eqref{BNPY} can be formulated as
		\begin{equation}
			\begin{gathered}
				\text { Find } \mathbf{R} \in \mathbb{C}^n, \mathbf{T} \in \mathbb{C}^{n+h} \\
				\text { s.t. } I(k, l)=\left|\mathbf{F}_{:, k}^{\mathrm{H}}\left(\mathbf{R} \odot\left(\mathbf{C}_l \mathbf{T}\right)\right)\right|^2+\varepsilon, \\
				k=1, \cdots, K, l=1, \cdots, L.\label{NNBPR}
			\end{gathered}
		\end{equation}
		The solution to Eq.~\eqref{NNBPR} is pursued using feasible methods that find a point in the intersection of multiple constraint sets. If noise $\varepsilon=0$, the ground truth is guaranteed to lie within this intersection. Therefore, the problem simplifies to finding a point that satisfies the constraints of two key sets:
		\begin{equation*}
			\begin{aligned}
				\mathcal{A} &:=\left\{ \boldsymbol{\Psi} := \left( \boldsymbol{\Psi}_1, \boldsymbol{\Psi}_2, \ldots, \boldsymbol{\Psi}_L \right) \in \mathbb{C}^{n \times L} \mid \right. \\
				& \quad \left. | \mathbf{F}_{:, k}^{\mathrm{H}} \boldsymbol{\Psi}_l |^2 = I(k, l), \, 1 \leq k \leq K, \, 1 \leq l \leq L \right\}, \\
				\mathcal{B} &:=\left\{ \boldsymbol{\Psi} := \left( \boldsymbol{\Psi}_1, \boldsymbol{\Psi}_2, \ldots, \boldsymbol{\Psi}_L \right) \in \mathbb{C}^{n \times L} \mid \exists~\mathbf{R}\in\mathbb{C}^n~\right. \\
				& \quad \text{and}~\mathbf{T} \in \mathbb{C}^{n+h}, \, \text{s.t.} \, \mathbf{R} \odot (\mathbf{C}_l\mathbf{T}) = \boldsymbol{\Psi}_l, \, 1 \leq l \leq L \left. \right\},
			\end{aligned}
		\end{equation*}
		where the elements in $\mathcal{A}$ satisfy the intensity constraints, and $\mathcal{B}$  requires the feasible set to contain the physical model and the information of the analyzer. Denote $\bm{\Psi}^*\in\mathcal{A}\cap\mathcal{B}$, then 
		\begin{eqnarray*}
			(\mathbf{R}^*,\mathbf{T}^*)&\in&\arg \min _{\mathbf{R}\in\mathbb{C}^n,\mathbf{T}\in\mathbb{C}^{n+h}} \frac{1}{2} \sum_{l=1}^L\left\|\mathbf{\Psi}_l^*-\mathbf{R} \odot (\mathbf{C}_l\mathbf{T})\right\|^2_2,
		\end{eqnarray*}
		which is the one of ground truth. Just like the gradient in the geometry-based methods, the definitions of projection on sets $\mathcal{A}$ and $\mathcal{B}$ namely $\mathbb{P}_{\mathcal{A}}$ and $\mathbb{P}_{\mathcal{B}}$ are also the foundation of feasible methods. Note that, $\mathcal{A}$ and $\mathcal{B}$ are closed sets, thus $\mathbb{P}_{\mathcal{A}}$ and $\mathbb{P}_{\mathcal{B}}$ exist and are derived as below respectively.
		
		Let $\Psi^{(m+1)_{\mathcal{A}}}:=\mathbb{P}_{\mathcal{A}}\left(\Psi^{(m)}\right)$, which can be estimated by solving the problem
		$$
		\mathbb{P}_{\mathcal{A}}\left(\boldsymbol{\Psi}^{(m)}\right)=\arg \min _{\boldsymbol{\Psi} \in \mathcal{A}} \sum_{l=1}^L\left\|\boldsymbol{\Psi}_l-\boldsymbol{\Psi}_l^{(m)}\right\|^2_2.
		$$
		More specifically, if $t_1, t_2, \cdots, t_K$ are multiples of $t_{\text {step }}$, and $n=\left\lfloor\frac{2 \pi}{t_{\text {step }} \Delta_{\text {step }}}\right\rfloor+1$ where $\lfloor\cdot\rfloor$ denotes rounding down, we have
		$$
		\begin{aligned}
			\mathbf{F}_{:, k} &= \frac{\Delta_{\text {step }}}{\sqrt{2 \pi}}\left(1, e^{i \Delta_{\text {step }} \cdot t_k}, \cdots, e^{i(n-1) \Delta_{\text {step }} \cdot t_k}\right) \\
			&\approx \frac{\Delta_{\text {step }}}{\sqrt{2 \pi}}\left(1, e^{\frac{2 \pi i\left\lfloor\frac{t_k}{t_{\text {step }}}\right\rfloor}{n}}, \cdots, e^{\frac{2 \pi i(n-1)\left\lfloor\frac{t_k}{t_{\text {step }}}\right\rfloor}{n}}\right) .
		\end{aligned}
		$$
		Next, define $\mathbf{F}_{n \times n}$ to be a standard Fourier matrix. Then, the $\left\lfloor\frac{t_k}{t_{\text {step }}}\right\rfloor$ th row of $\frac{\Delta_{\text {step }}}{\sqrt{2 \pi}} \mathbf{F}_{n \times n}$ is $\mathbf{F}_{:, k}^{\mathrm{H}}$, where $(\cdot)^{\mathrm{H}}$ denotes the conjugate transpose. Next, let $\hat{\boldsymbol{\Psi}}_l^{(m+1)}=\frac{\Delta_{\text {step }}}{\sqrt{2 \pi}} \mathbf{F}_{n \times n} \boldsymbol{\Psi}_l^{(m)}, l= 1,2, \cdots, L$. Introducing $\mathcal{I}=\left\{\left\lfloor\frac{t_1}{t_{\text {step }}}\right\rfloor,\left\lfloor\frac{t_2}{t_{\text {step }}}\right\rfloor, \cdots,\left\lfloor\frac{t_K}{t_{\text {step }}}\right\rfloor\right\}$, we modify $\hat{\boldsymbol{\Psi}}_l^{(m+1)}$ as
		
		$$
		\hat{\Psi}_l^{(m+1)}(k)=\left\{\begin{array}{l}
			\sqrt{I_l(k)} \odot \frac{\hat{\Psi}_l^{(m)}(k)}{\left|\hat{\Psi}_l^{(m)}(k)\right|}, k \in \mathcal{I} \\
			\hat{\Psi}_l^{(m)}(k), \text { else }
		\end{array},\right.
		$$ where $l=1,2,\cdots,L$.
		Finally, we can estimate $\mathbb{P}_{\mathcal{A}}\left(\boldsymbol{\Psi}^{(m)}\right)$ as below
		$$
		\begin{aligned}
			\mathbb{P}_{\mathcal{A}}\left(\mathbf{\Psi}^{(m)}\right) &= \left\{ \frac{\sqrt{2 \pi}}{\Delta_{\text {step }}} \mathbf{F}_{n \times n}^{-1} \hat{\boldsymbol{\Psi}}_1^{(m+1)}, \frac{\sqrt{2 \pi}}{\Delta_{\text {step }}} \mathbf{F}_{n \times n}^{-1} \hat{\boldsymbol{\Psi}}_2^{(m+1)},\right. \\
			&\quad \left.  \cdots, \frac{\sqrt{2 \pi}}{\Delta_{\text {step }}} \mathbf{F}_{n \times n}^{-1} \hat{\boldsymbol{\Psi}}_L^{(m+1)} \right\}.
		\end{aligned}
		$$
		The projection $\boldsymbol{\Psi}^{(m+1)_{\mathcal{B}}}:=\mathbb{P}_{\mathcal{B}}\left(\boldsymbol{\Psi}^{(m)}\right)$ can be calculated via the following two steps:
		$$
		\begin{aligned}
			&\text{i)} \quad \left(\mathbf{R}^{(m+1)}, \mathbf{T}^{(m+1)}\right) = \arg \min _{\mathbf{R}\in\mathbb{C}^n, \mathbf{T}\in\mathbb{C}^{n+h}} \\
			&\quad \qquad\qquad\qquad\qquad\qquad \frac{1}{2} \sum_{l=1}^L\left\|\bm{\Psi}_l^{(m)}-\mathbf{R} \odot\left(\mathbf{C}_l \mathbf{T}\right)\right\|^2_2, \\
			&\text{ii)} \quad \mathbf{\Psi}_l^{(m+1)_{\mathcal{B}}} = \mathbf{R}^{(m+1)} \odot\left(\mathbf{C}_l \mathbf{T}^{(m+1)}\right),l=1,2, \cdots, L.
		\end{aligned}
		$$
		In practice, $\textrm{i})$ can be solved inexactly by the alternating minimization method as
		$$
		\begin{aligned}
			\mathbf{R}^{(m+1)} & =\arg \min _{\mathbf{R} \in \mathbb{C}^n} \frac{1}{2} \sum_{l=1}^L\left\|\mathbf{\Psi}_l^{(m)}-\mathbf{R} \odot\left(\mathbf{C}_l \mathbf{T}^{(m)}\right)\right\|^2_2 \\
			& =\frac{\sum_{l=1}^L \overline{\left(\mathbf{C}_l \mathbf{T}^{(m)}\right)} \odot \mathbf{\Psi}_l^{(m)}}{\sum_{l=1}^L\left|\mathbf{C}_l \mathbf{T}^{(m)}\right|^2},
		\end{aligned}
		$$
		where $\overline{~\cdot~}$, $|\cdot|^2$, and $\frac{~\cdot~}{~\cdot~}$ are the element-wise operators, and
		\begin{eqnarray*}
			\mathbf{T}^{(m+1)}=\arg\min_{\mathbf{T}
				\in\mathbb{C}^{n+h}}\frac{1}{2} \sum_{l=1}^L\left\|\mathbf{\Psi}_l^{(m)}-\mathbf{R}^{(m+1)} \odot\left(\mathbf{C}_l\mathbf{T}\right)\right\|^2_2. 
		\end{eqnarray*} 
		If the matrix $\sum_{l=1}^L \mathbf{C}_l^{\mathrm{H}} \operatorname{diag}\left(\left|\mathbf{R}^{(m+1)}\right|^2\right)_{n \times n} \mathbf{C}_l$ is invertible, the solution to optimization problem above has a closed form
		$$
		\begin{aligned}
			\mathbf{T}^{(m+1)} &= \left(\sum_{l=1}^L \mathbf{C}_l^{\mathrm{H}} \operatorname{diag}\left(\left|\mathbf{R}^{(m+1)}\right|^2\right)_{n \times n} \mathbf{C}_l\right)^{-1} \\
			&\quad \left(\sum_{l=1}^L \mathbf{C}_l^{\mathrm{H}}\left(\overline{\mathbf{R}^{(m+1)}} \odot \mathbf{\Psi}_l^{(m)}\right)\right),
		\end{aligned}
		$$
		where $\operatorname{diag}(\cdot)_{n \times n}$ is to transform a length $n$ vector into an $n \times n$ diagonal matrix. Else, the gradient descent algorithm can be applied to estimate the solution.

		The problem therefore reduces to the development of an algorithm based on a careful synthesis of the projection operators $\mathbb{P}_{\mathcal{A}}(\cdot)$ and $\mathbb{P}_{\mathcal{B}}(\cdot)$ to locate a point within the intersection of $\mathcal{A}$ and $\mathcal{B}$. We investigate three alternative methods for this purpose
		\begin{itemize}
			\item The Alternating Projections (AP) method: A classical projection-based algorithm.
			\item The Douglas-Rachford (DR) method: An algorithm based on applying reflections.
			\item The Relaxed Averaged Alternating Reflections (RAAR) method: A generalized approach that can be viewed as a combination of the AP and DR schemes.
		\end{itemize}
		A comparative summary of these methods is provided in Table \ref{TB}. We apply these algorithms to reconstruct the response of the previously discussed $\alpha$-Fe target from the numerical 2D spectrum, without relying on any prior knowledge of the analyzer. The respective results are shown in Fig.~\ref{Fig.4}.
		
		First, we conduct comparative tests to demonstrate the superiority of feasible method over L-BFGS method, and the results are displayed in Fig.~\ref{Fig.4}(a)-(e). Here, we use the L-BFGS method to solve the following equation 
		\begin{equation}
			\min_{\mathbf{R},\mathbf{T}}\ell(\mathbf{R},\mathbf{T})=\frac{1}{2KL}\sum\limits_{k=1}^{K}\sum\limits_{l=1}^{L}\left(|\Psi(k,l)|-\sqrt{I(k,l)}\right)^2\label{landscape},
		\end{equation}
		where $\Psi(k,l) = \mathbf{F}^{\text{H}}_{:,k}(\mathbf{R}\odot(\mathbf{C}_l\mathbf{T}))$. Notice that the variable of interest is the joint vector $\tilde{\mathbf{R}} = [\mathbf{R}; \mathbf{T}] \in \mathbb{C}^{2n+h}$, which concatenates the sample $\mathbf{R}$ and the analyzer $\mathbf{T}$. Consequently, the gradient $\nabla\ell(\tilde{\mathbf{R}})=[\frac{\partial\ell(\tilde{\mathbf{R}})}{\partial\mathbf{R}};\frac{\partial\ell(\tilde{\mathbf{R}})}{\partial\mathbf{T}}]\in\mathbb{C}^{2n+h}$ is also computed accordingly using its respective partial derivative formulations. 
		
		In contrast to its behavior in Fig.~\ref{Fig.2} and \ref{Fig.3}, the L-BFGS method exhibits deteriorated performance in this test. This is evidenced by oscillations in the recovered spectrum shown in Fig.~\ref{Fig.4}(a) and a tangential convergence trajectory relative to the contours of relative error displayed in Fig.~\ref{Fig.4}(d), indicating convergence to a stagnation point (yellow diamond). This point, located within a basin near the ground truth in the loss landscape of Eq.~\eqref{landscape} shown in Fig.~\ref{Fig.4}(b), is characterized geometrically by its $\ell_2$ norm of gradient and the extreme eigenvalues of Hessian, which identifies it near a degenerate saddle point. The geometry-based optimization methods usually get trapped in the vicinity of this saddle point \cite{jin2017escape}. 
		
		Subsequently, the RAAR algorithm successfully enables the iterative sequence to escape this local basin of attraction, which is evidenced in Fig.~\ref{Fig.4}(d) and (e), where both the relative error and the measurement error resume a decreasing trend. 
		Notably, within the region marked by the yellow dashed box—corresponding to the boundary between Phase I and Phase II in Fig.~\ref{Fig.4} (e) shows that the measurement error does not decrease monotonically, which illustrates the RAAR method's process of pulling the solution out of stagnation vividly. When getting stagnated, the measurement error of RAAR is further reduced by adjusting the parameter $\alpha$, as seen in the region within the red dashed box in Fig.~\ref{Fig.4}(e). Notably, $\alpha$ plays a critical role in balancing the projector and reflector for RAAR method, as formulated in Table \ref{TB}. By tuning parameter $\alpha$, the RAAR method can maintain dynamic and prevent stagnation into local minimum earlier. Finally, the estimated intensities of $\mathbf{R}$ and $\mathbf{T}$ by RAAR, shown in Fig.~\ref{Fig.4}(c), fits the ground truth better with significantly fewer oscillations than those recovered by the L-BFGS method. This result fully verifies the superiority of feasible method such as RAAR over the geometry-based method when dealing with blind Nuclear Ptychoscopy problem.
		
		Notably, the 1D Nuclear Ptychoscopy setup in this work offers low computational overhead, enabling detailed geometric analysis (e.g., loss landscape visualization, saddle point characterization) that is often intractable with 2D ptychography. This 1D-derived insight clarifies why feasible methods are powerful for higher-dimensional blind ptychography: 2D problems exhibit more severe non-convexity, making the stagnation issue of geometry-based methods more pronounced\cite{chang2022fast}.
		
		Specifically, to maintain the non-convex structure of the landscape when embedding it into 3D space, the directions are chosen elaborately. In Fig.~\ref{Fig.4}(b), one of the directions $\mathbf{d}_1\in\mathbb{C}^{2n+h}$ for plotting the landscape is the difference between the initialization point and the stagnation point of the L-BFGS method. Another direction $\mathbf{d}_2\in\mathbb{C}^{2n+h}$ is the difference between the final result of RAAR method and the stagnation point of L-BFGS method. Then, every point in Fig.~\ref{Fig.4}(b) is determined by $$\left(\beta_1,\beta_2,\ell(\beta_1\frac{\bm{d}_1}{\|\bm{d}_1\|_2}+\beta_2\frac{\bm{d}_2}{\|\bm{d}_2\|_2})\right),$$
		where $\beta_1$ and $\beta_2$ are the coordinates of x-axis and y-axis. $\ell(\cdot)$ is the loss function shown in Eq.~\eqref{landscape}. Here, $\ell(\alpha\frac{\bm{d}_1}{\|\bm{d}_1\|_2}+\beta\frac{\bm{d}_1}{\|\bm{d}_1\|_2})$ is the simplification of
		\begin{eqnarray*}
			\ell\left(\left(\beta_1\frac{\bm{d}_1}{\|\bm{d}_1\|_2}+\beta_2\frac{\bm{d}_2}{\|\bm{d}_2\|_2}\right)_{1:n},\right.\\
			\left.\left(\beta_1\frac{\bm{d}_1}{\|\bm{d}_1\|_2}+\beta_2\frac{\bm{d}_2}{\|\bm{d}_2\|_2}\right)_{n+1:2n+h}\right)
		\end{eqnarray*}
		where $(\cdot)_{l_1:l_2}$ is the operator that extracts the elements located from the $l_1$th to the $l_2$th position of the vector.
		
		Meanwhile, for the trajectories shown in Fig.~\ref{Fig.4}(d), the two directions are estimated via principal component analysis (PCA) to a matrix whose columns are the displacement vectors from the final RAAR solution to all algorithmic (L-BFGS and RAAR) iterates \cite{jolliffe2011principal}, and the coordinates of the trajectories are determined by the projections on these directions.
		
		Usually, the explicit construction of the Hessian $\nabla^2\ell(\mathbf{R},\mathbf{T})$ of Eq.~\eqref{landscape} is complicated, especially when the length of the variables $\mathbf{R}$ and $\mathbf{T}$ is large. Specifically, the dimension of the Hessian matrix is about $16000\times16000$ in the tests. So, it is computationally costly to directly evaluate its eigenvalues for such a large matrix. Instead, we utilize the implicitly restarted Lanczos method to estimate the extreme eigenvalues \cite{lanczos1950iteration} shown in Fig.~\ref{Fig.4}(b), which only needs to construct the product between the Hessian matrix and a vector. More details about landscape visualization, PCA of trajectories and Hessian spectral analysis can be found in \cite{li2018visualizing}.
		
		After the comparative test, we next evaluate the performance of three feasible methods: AP, DR, and RAAR algorithms. For the RAAR method, a parameter tuning strategy was implemented, with $\alpha=1$ for the first $21000$ iterations and $\alpha=0.98$ thereafter. The results are presented in Fig.~\ref{Fig.4}(f-k).
		
		As shown in Fig.~\ref{Fig.4}(g), the measurement error of the AP method decreases rapidly in the initial phase. However, the oscillatory behavior in the recovered intensities of $\mathbf{R}$ and $\mathbf{T}$ in Fig.~\ref{Fig.4}(h) and (j) indicate that it stagnates at suboptimal solutions. This is further verified by the relative error metrics in Fig.~\ref{Fig.4}(i) and (k), highlighting that AP is susceptible to stagnation when applied to non-convex problems.
		
		
		In contrast to AP, which relies solely on direct projections, both DR and RAAR incorporate reflection operators. These approaches have been empirically demonstrated to enhance the ability to escape from local minima in non-convex optimization \cite{yuan2024phase,elser2003phase}.
		Notably, Table \ref{TB} indicates that DR is equivalent to RAAR with $\alpha=1$. Consequently, their measurement and relative error curves are identical for the first $21000$ iterations, as seen in Fig.~\ref{Fig.4}(g), (i), and (k). Visually, the intensities and phases of $\mathbf{R}$ and $\mathbf{T}$ recovered by both methods, shown in Fig.~\ref{Fig.4}(h) and (j), align well with the ground truth.
		
		A critical distinction emerges after the $21000$th iteration indicated by the dashed grey line in Fig.~\ref{Fig.4}(g). While the DR method itself stagnates, the parameter tuning strategy adopted by the RAAR method reactivates the convergence process, allowing RAAR to achieve a lower final measurement error and final relative error for both $\mathbf{R}$ and $\mathbf{T}$. Therefore, among the feasible methods investigated, RAAR is preferable for dealing with the blind Nuclear Ptychoscopy problem.
		
		\subsection{Constrained optimization methods}
		We have so far considered two scenarios: reconstruction with a known analyzer response and blind reconstruction, in which no prior information about the analyzer or the target is available. In this section, we turn to the case where partial prior information about the target, the analyzer, or both is available. Leveraging such prior knowledge is a well-established strategy in inverse problems—including super-resolution \cite{shi2015lrtv}, deblurring \cite{beck2009fast}, and hyperspectral imaging \cite{yuan2012hyperspectral}, where it mitigates ill-posedness and enhances stability, especially in the presence of measurement noise. Extending this principle to Nuclear Ptychoscopy, we demonstrate how prior information can be incorporated to enhance reconstruction performance, with particular attention to robustness under different noise levels.
		
		To formalize the integration of prior knowledge, we refer to Eq.~\eqref{bayesian} (the Bayesian formulation), where prior information of $\mathbf{R}$ can be incorporated as a regularization term to guide the optimization. We first consider the case where the analyzer response is known and partial prior information about the target is available—for example, assuming $\mathbf{R}$ is smooth, which is a very loose constraint. Next, we consider the situation where the analyzer response is unknown, but the analyzer’s time spectrum is available as a constraint. Finally, we combine these two conditions into a comprehensive scenario: the analyzer response is unknown (but its time spectrum is constrained), and a smooth prior on $\mathbf{R}$ is imposed.
		\subsubsection{Smooth Target Response Prior with Known Analyzer}
		Given that the analyzer response is known, the geometry-based method is favored. This leads to an optimization formulation that integrates the prior on $\mathbf{R}$ as below
		\begin{eqnarray}\label{Universal Prior}
			\min_{\mathbf{R}\in\mathbb{C}^n}\ell(\mathbf{R})+\tau\mathcal{R}(\mathbf{R}),
		\end{eqnarray}
		where $\ell(\mathbf{R})$ is the loss function defined previously, $\tau\geq0$ is a regularization parameter that balances the contributions of the loss and the regularizer, and $\mathcal{R}(\cdot)$ is a regularizer designed to encode specific prior knowledge. For instance, employing a TV regularizer, defined for a sequence as:
		\begin{eqnarray}\label{TV}   
			\min_{\mathbf{R}\in\mathbb{C}^n} \|\mathbf{R}\|_{\textrm{TV}} = \sum_{i=1}^{n-1} \Big| \, |\mathbf{R}(\Delta_i)| - |\mathbf{R}(\Delta_{i-1})| \, \Big|, 
		\end{eqnarray}
		promotes piecewise smoothness by penalizing large fluctuations between adjacent elements.
		Since $\mathcal{R}(\mathbf{R})$ such as TV regularizer is often non-differentiable, the geometry-based method cannot be applied directly to solve Eq.~\eqref{Universal Prior}. 
		
		A common strategy to address this is to decouple the loss and regularization terms using a proximal optimization approach. At iteration $m$, the proximal gradient descent algorithm proceeds in two steps:
		\begin{itemize}
			\item \textbf{Gradient Descent Step}: Update the variable by moving against the gradient of the loss function:
			$$\mathbf{R}^{(m+\frac{1}{2})}=\mathbf{R}^{(m)}-\lambda\nabla\ell(\mathbf{R}^{(m)}).$$
			\item \textbf{Proximal Mapping Step}: Apply the proximal operator of the regularizer $\mathrm{prox}_{\lambda\tau\mathcal{R}}(\cdot)$ to the intermediate result 
			$\mathbf{R}^{(m+\frac{1}{2})}$:
			$$
			\begin{aligned}
				\mathbf{R}^{(m+1)} &= \mathrm{prox}_{\lambda\tau\mathcal{R}}(\mathbf{R}^{(m+\frac{1}{2})}) \\
				&= \arg\min_{\mathbf{R}\in\mathbb{C}^n} \mathcal{R}(\mathbf{R}) + \frac{1}{2\tau\lambda}\|\mathbf{R}-\mathbf{R}^{(m+\frac{1}{2})}\|^2_2.
			\end{aligned}
			$$
			
		\end{itemize}
		Intuitively, the proximal mapping step minimizes the regularizer $\mathcal{R}(\mathbf{R})$ while ensuring that the updated variable $\mathbf{R}^{(m+1)}$ remains in close proximity to the gradient-based update $\mathbf{R}^{(m+\frac{1}{2})}$, thus enforcing the prior without diverging from the data fidelity objective.
		
		An alternative approach is to employ variable splitting via the Alternating Direction Method of Multipliers (ADMM). This method introduces an auxiliary variable $\mathbf{Y} \in \mathbb{C}^n$ and a dual variable $\mathbf{Z} \in \mathbb{C}^n$ to decouple the terms $\ell(\mathbf{R})$ and $\mathcal{R}(\mathbf{R})$. The augmented Lagrangian for problem Eq.~\eqref{Universal Prior} is formulated as follows:
		\begin{equation}
			\begin{split}
				\min_{\mathbf{R}\in\mathbb{C}^n,\mathbf{Y}\in\mathbb{C}^n,\mathbf{Z}\in\mathbb{C}^n} &\ell(\mathbf{R})+\frac{\eta}{2}\|\mathbf{R}-\mathbf{Y}\|^2_2 \\
				&+\operatorname{Re}\left( \langle \mathbf{Z}, \mathbf{R} - \mathbf{Y} \rangle \right)+\tau\mathcal{R}(\mathbf{Y}),
			\end{split}
		\end{equation}
		where $\eta\geq0$ is a penalty parameter that controls the consensus between $\mathbf{R}$ and $\mathbf{Y}$. The dual variable $\mathbf{Z}$ acts as a Lagrange multiplier, reducing the algorithm's sensitivity to the choice of $\eta$ and ensuring that the estimates for $\mathbf{R}$ and $\mathbf{Y}$ converge to a consensus even for moderate values of the penalty parameter.
		
		The ADMM algorithm iteratively minimizes this Lagrangian by solving three sequential subproblems at each iteration $m$:
		\begin{equation*}
			\left\{    
			\begin{aligned}
				&\mathbf{R}^{(m+1)}\in\arg\min_{\mathbf{R}\in\mathbb{C}^n}\ell(\mathbf{R})+\frac{\eta}{2}\|\mathbf{R}-\mathbf{Y}^{(m)}+\mathbf{Z}^{(m)}\|^2_2\\
				&\mathbf{Y}^{(m+1)}\in\arg\min_{\mathbf{Y}\in\mathbb{C}^n}\frac{\eta}{2}\|\mathbf{Y}-\mathbf{R}^{(m+1)}-\mathbf{Z}^{(m)}\|^2_2+\tau\mathcal{R}(\mathbf{Y})\\
				&\mathbf{Z}^{(m+1)} = \mathbf{Z}^{(m)}+\lambda(\mathbf{R}^{(m+1)}-\mathbf{Y}^{(m+1)})			\end{aligned}\right.,
		\end{equation*}
		where $\lambda$ denotes the learning rate. For the traditional ADMM method, $\lambda$ is theoretically set to $\eta$. Nevertheless, this parameter allows for adaptive tuning in practice.
		\begin{figure*}[ht]
			\setlength{\abovecaptionskip}{-0.2cm} 
			\centerline{\includegraphics[width=1.00\linewidth]{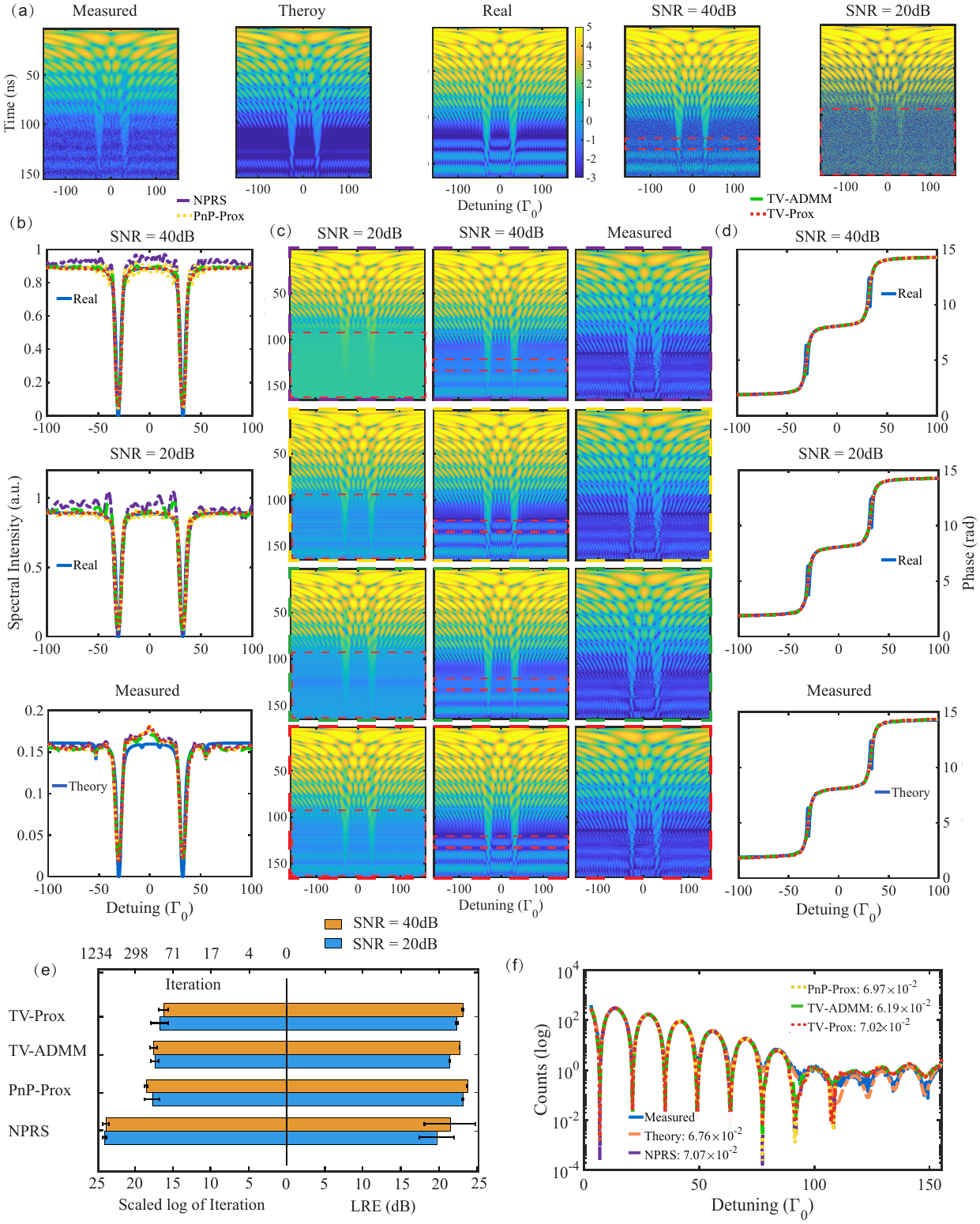}}
			\centering
			\begin{picture}(10,5)
			\end{picture}
			\caption{Numerical and experimental evaluation of constrained optimization methods with a smooth target prior, showing reconstruction performance under varying noise conditions. (a) Simulated and experimental input data. Top row: the true 2D spectrum and simulated measurements corrupted by noise at SNRs of 40 dB and 20 dB. Bottom row: the experimentally measured 2D spectrum and its theoretical estimation. (b) Recovered intensity, (d) recovered phase, and (c) reconstructed 2D spectrum obtained using NPRS, PnP-Prox, TV-ADMM, and TV-Prox methods for simulated data under varying noise levels and for experimental data. (e) Quantitative comparison of different methods based on recovered logarithmic relative error (LRE) and number of iterations using histograms. The heights of the histograms represent the mean values, and the corresponding error bars stand for the variance. (f) Comparison of time spectrum calculated from the recovered amplitude and phase for each method, alongside the theoretical fit and experimental data, and the measurement errors are also calculated.}
			\label{Fig.5} 
		\end{figure*}
		
		Notice that, both the proximal gradient descent and ADMM method solve a common subproblem simplified as below,
		\begin{eqnarray}
			\min_{\mathbf{R}\in\mathbb{R}^n}\frac{1}{2}\|\mathbf{R}-\mathbf{V}\|^2_2+\tau\mathcal{R}(\mathbf{R}),\label{PnP}
		\end{eqnarray}
		where $\mathbf{V}$ is an intermediate variable. When $\mathcal{R}(\cdot)$ is the TV prior, the solution of Eq.~\eqref{PnP} can be estimated by many efficient solvers \cite{condat2013direct}. Rather than explicitly defining and minimizing $\mathcal{R}(\cdot)$, PnP methods treat the solution of Eq.~\eqref{PnP} as a denoising operation to get a smooth solution. The idea behind this is that Eq.~\eqref{PnP} is to balance `keeping $\mathbf{R}$ close to $\mathbf{V}$' and `making $\mathbf{R}$ smooth via $\mathcal{R}(\cdot)$'. Consequently, since state-of-the-art denoisers inherently encode the smoothness constraint of 
		$\mathcal{R}(\cdot)$, the minimization can be approximately performed by applying an off-the-shelf Gaussian noise denoiser, such as a wavelet denoiser, to the vector $\mathbf{V}$ \cite{venkatakrishnan2013plug}.
		
		Compared to the TV regularizer, the PnP approach equipped with a wavelet denoiser is often considered more user-friendly as it avoids the need for manual tuning of the functional form of $\mathcal{R}(\cdot)$ and can be simpler to implement. Here, we integrate the PnP method into the proximal gradient descent framework (denoted as PnP-Prox) and compared its performance to the following algorithms: NPRS, TV-ADMM (ADMM regularized by TV), and TV-Prox (proximal gradient descent regularized by TV). 
		
		As mentioned, incorporating partial prior information has been shown to improve reconstruction performance in ptychography, particularly under noisy measurement conditions \cite{chang2018total}. To evaluate the benefit of this approach in Nuclear Ptychoscopy, we apply the NPRS, PnP-Prox, TV-ADMM, and TV-Prox methods to the $\alpha$-Fe case considered in the previous section, using both a noise-corrupted numerical and experimental 2D spectrum. The corresponding results are presented in Fig.~\ref{Fig.5}. The tests consist of two parts:
		\begin{itemize}
			\item Simulation: The measurement vector $\mathbf{I}$ was corrupted with additive white Gaussian noise at signal-to-noise ratios (SNR)\footnote{Define $\mathbf{x}\in\mathbb{R}^n$ (or $\mathbb{C}^n$ if complex) to be the real (or complex) corrupted by the additive noise $\mathbf{v}\in\mathbb{R}^n$(or $\mathbb{C}^n$). Then, the signal to noise ratio (SNR) of $\tilde{\mathbf{x}}=\mathbf{x}+\mathbf{v}$ is defined as: $$\text{SNR}:=20\log_{10}\frac{\|\mathbf{x}\|_2}{\|\mathbf{v}\|_2}.$$} of 20 dB and 40 dB. To establish statistical reliability, each condition was tested over $20$ independent trials shown in Fig.~\ref{Fig.5}(e).
			\item Experiment: The methods were applied to the experimental dataset previously introduced in Fig.~\ref{Fig.3}. The performance of the algorithms is verified by the fitness to the independent time measurement of $\mathbf{R}$ shown in Fig.~\ref{Fig.5}(f).
		\end{itemize}
		At the same time, the logarithmic relative error (LRE) metric provides a convenient way to visualize reconstruction accuracy via histograms. Its definition for the recovered response $\tilde{\mathbf{R}}$ parallels that of the conventional signal-to-noise ratio (SNR), as both quantify logarithmic ratios of a reference quantity to an error term. Specifically,
		\begin{eqnarray}
			\text{LRE}&=&-20\log_{10}\left(\text{Relative Error}\right)\nonumber\\
			&=&20\log_{10}\frac{\|\mathbf{R}\|_2}{\|\tilde{\mathbf{R}}-\mathbf{R}\|_2},\label{LRE}
		\end{eqnarray}
		where $\mathbf{R}$ denotes the ground truth, and the difference $\tilde{\mathbf{R}}-\mathbf{R}$ is treated as the effective noise in the computation. Unlike SNR, which is typically defined for measurement data, the LRE directly quantifies reconstruction accuracy relative to the ground truth. A higher LRE thus indicates closer agreement with the ground truth and correspondingly improved reconstruction quality.
		
		The results are displayed in Fig.~\ref{Fig.5}. In simulation tests under varying noise levels, incorporating prior information significantly improves reconstruction quality. As shown in Fig.~\ref{Fig.5}(b), the intensities recovered by the PnP-Prox, TV-ADMM, and TV-Prox methods more closely match the ground truth than those obtained with NPRS, exhibiting fewer oscillations. Simultaneously, the phases recovered by all methods, shown in Fig.~\ref{Fig.5}(d), are in good agreement with the true values. This improvement is further quantified by the histogram in Fig.~\ref{Fig.5}(e), which shows that the mean LRE of the estimates $\tilde{\mathbf{R}}$ is highest for PnP-Prox, with a small variance, followed by TV-Prox, and then TV-ADMM and the NPRS method. The mean number of iterations required to achieve the corresponding LRE was also recorded. While the NPRS method required over $1000$ iterations, the other three methods converged in approximately $200$ iterations. 
		
		
		In Fig.~\ref{Fig.5}(a), the feature within the orange dashed box in the 2D time-energy spectrum (SNR = 40 dB) is indistinct. However, this feature is successfully reconstructed in the results presented in  Fig.~\ref{Fig.5}(c), particularly by PnP-Prox and TV-Prox methods. Notably, even at an SNR of $20\text{dB}$, where the measurements in Fig.~\ref{Fig.5}(a) are severely corrupted, the features in the orange dashed boxed region of Fig.~\ref{Fig.5}(c) can also be recovered by the PnP-Prox, TV-ADMM and TV-Prox methods, whereas the NPRS method fails. These results clearly illustrate the significant advantage of incorporating prior information into reconstruction algorithms when dealing with noisy measurements.
		
		
		For experimental data, as shown in Fig.~\ref{Fig.5}(b), (c) and (d), all methods performed competently, recovering the intensity, phase, and 2D spectrum in good agreement with the theoretical results. However, as shown in Fig.~\ref{Fig.5}(f), the relative errors of the time spectrum for $\mathbf{R}$ recovered by the TV-Prox, TV-ADMM, and PnP-Prox methods are lower than that of the NPRS method. In particular, the result from the PnP-Prox method exhibits a lower error than even the theoretical benchmark. This underscores the practical advantage of incorporating prior information in experimental Nuclear Ptychoscopy.
		
		\subsubsection{Smooth Target Response Prior and Analyzer Time-Spectrum Constraint}
		We have discussed incorporating prior information about the target when the analyzer response is known. We now extend the discussion to the case where the analyzer response is unknown, but its time spectrum is available as a constraint, referred to as a measurement prior in the following. In this case, as previously noted, feasible methods provide an effective framework. The constrained blind Nuclear Ptychoscopy problem can then be formulated as
		\begin{eqnarray}\label{RAARcon}&\min\limits_{\mathbf{R}\in\mathbf{C}^n,\mathbf{T}\in\mathbb{C}^{n+h}}\mathcal{R}_1(\mathbf{R})+\mathcal{R}_2(\mathbf{T})\nonumber\\
			&\mathrm{s.t.}~\bm{\Psi}\in\mathcal{A}\cap\mathcal{B},
		\end{eqnarray}
		where $\bm{\Psi}=\{\mathbf{R}\odot(\mathbf{C}_1\mathbf{T}),\mathbf{R}\odot(\mathbf{C}_2\mathbf{T}),\cdots,\mathbf{R}\odot(\mathbf{C}_L\mathbf{T})\}$, with $\mathcal{R}_1(\cdot)$ and $\mathcal{R}_2(\cdot)$ acting as regularizers for the sample and analyzer, respectively, and $\mathbf{C}_l$ being the shifting matrix defined earlier. $\mathcal{A}$ and $\mathbf{B}$ are two feasible sets introduce above. The problem described by Eq.~\eqref{RAARcon} is addressed via a constrained RAAR algorithm. At the $m$-th iteration, the algorithm proceeds as follows:
		\begin{itemize}
			\item \textbf{RAAR Update}: Apply the standard RAAR projection shown in Tab.~\ref{TB} to $\bm{\Psi}^{(m)}$ to compute $\bm{\Psi}^{(m+1)}$.
			\item 
			\textbf{Regularized Reconstruction}: Update the sample and analyzer by solving the regularized least-squares problems: 
			
			
			\begin{equation*}
				\left\{
				\begin{aligned}
					\mathbf{R}^{(m+1)} &= \arg \min \limits_{\mathbf{R}\in\mathbb{C}^n} \frac{1}{2} \sum_{l=1}^L\left\|\mathbf{\Psi}_l^{(m+1)}-\mathbf{R} \odot\left(\mathbf{C}_l \mathbf{T}^{(m)}\right)\right\|^2_2 \\
					& \quad +\tau_1\mathcal{R}_1(\mathbf{R}), \\
					\mathbf{T}^{(m+1)} &= \arg \min \limits_{\mathbf{T}\in\mathbb{C}^{n+h}} \tau_2\mathcal{R}_2(\mathbf{T}) \\
					& \quad +\frac{1}{2} \sum_{l=1}^L\left\|\mathbf{\Psi}_l^{(m+1)}-\mathbf{R}^{(m+1)}   \odot\left(\mathbf{C}_l \mathbf{T}\right)\right\|^2_2,
				\end{aligned}
				\right.
			\end{equation*}
			with regularization parameters $\tau_1$ and $\tau_2$ controlling the strength of the penalties imposed by $\mathcal{R}_1(\cdot)$ and $\mathcal{R}_2(\cdot)$, respectively.
		\end{itemize}
		Here, $\mathcal{R}_1(\cdot)$ and $\mathcal{R}_2(\cdot)$ can also use the TV regularizer or the two constrained least square problems can also be solved by using the PnP method discussed in previous section. Furthermore, often a time spectrum of the analyzer alone $$I_{\mathbf{T}}(k)=\left|\mathbf{F}_{:, k}^{\mathrm{H}} \mathbf{C}_{\mathbf{T}} \mathbf{T}\right|^2+\varepsilon_2, k=1, \cdots, K,$$ is available in experiments, where
		
		\begin{equation*}
			\setlength{\arraycolsep}{2pt} 
			\renewcommand{\arraystretch}{1} 
			\mathbf{C}_{\mathbf{T}} = 
			\left(
			\begin{array}{c|cccccccccc}
				& \textcolor{blue}{1} & \textcolor{blue}{2} &\textcolor{blue}{\dots} & \textcolor{blue}{f_{\mathbf{T}}+1} & \textcolor{blue}{f_{\mathbf{T}}+2} & \textcolor{blue}{\dots} & \textcolor{blue}{f_{\mathbf{T}}+n} & \textcolor{blue}{\dots} &\textcolor{blue}{ h+n} \\
				\hline
				\textcolor{blue}{1} & 0 & 0 & \dots & 1 & 0 & \dots & 0 & \dots & 0 \\
				\textcolor{blue}{2} & 0 & 0 & \dots & 0 & 1 & \dots & 0 & \dots & 0 \\
				\textcolor{blue}{\vdots} & \vdots & \vdots & \dots & \vdots & \vdots & \ddots & \vdots & \dots & \vdots \\
				\textcolor{blue}{n} & 0 & 0 & \dots & 0 & 0 & \dots & 1 & \dots & 0
			\end{array}
			\right),
		\end{equation*}
		and $f_{{\mathbf{T}}}=\frac{\Delta^{1^+}_D}{\Delta_{\text{step}}}$. Then feasible set $\mathcal{A}$ and $\mathcal{B}$ have the formulation below:
		$$
		\begin{aligned}
			\mathcal{A} & := \left\{ \boldsymbol{\Psi} := \left( \boldsymbol{\Psi}_1, \boldsymbol{\Psi}_2, \ldots, \boldsymbol{\Psi}_L, \boldsymbol{\Psi}_{\mathbf{T}} \right) \in \mathbb{C}^{n \times(L+1)} \mid \right. \\
			& \quad \left| \mathbf{F}_{:, k}^{\mathrm{H}} \boldsymbol{\Psi}_l \right|^2 = I(k, l), \, 1 \leq k \leq K, \, 1 \leq l \leq L, \\
			& \quad \left. \left| \mathbf{F}_{:, k}^{\mathrm{H}} \boldsymbol{\Psi}_{\mathbf{T}} \right|^2 = I_{\mathbf{T}}(k), \, 1 \leq k \leq K \right\}, \\
			\mathcal{B} & := \left\{ \boldsymbol{\Psi} := \left( \boldsymbol{\Psi}_1, \boldsymbol{\Psi}_2, \ldots, \boldsymbol{\Psi}_L, \boldsymbol{\Psi}_{\mathbf{T}} \right) \in \mathbb{C}^{n \times(L+1)} \mid \right. \\
			& \quad \exists~\mathbf{R} \in \mathbb{C}^n, \, \mathbf{T} \in \mathbb{C}^{n+h}, \, \text{s.t.} \, \mathbf{C}_{\mathbf{T}} \mathbf{T} = \boldsymbol{\Psi}_{\mathbf{T}}, \\
			& \quad \left. \mathbf{R} \odot \left( \mathbf{C}_l \mathbf{T} \right) = \boldsymbol{\Psi}_l, \, 1 \leq l \leq L \right\}.
		\end{aligned}
		$$
		The constraints $\mathcal{A}$ and $\mathcal{B}$ are then incorporated into the RAAR framework. Note that the procedure to estimate $\mathbf{R}^{(m+1)}$ is identical to that in the standard RAAR method listed in Tab.\ref{TB}, and the remaining subproblem to address at each iteration is the update of the analyzer:
		$$
		\begin{aligned}
			\mathbf{T}^{(m+1)} &= \arg \min_{\mathbf{T}} \frac{1}{2} \sum_{l=1}^L\left\|\Psi_l^{(m+1)} - \mathbf{R}^{(m+1)} \odot\left(\mathbf{C}_l \mathbf{T}\right)\right\|^2_2 \\
			& \quad + \tau_2\left\|\Psi_{\mathbf{T}}^{(m+1)} - \mathbf{C}_{\mathbf{T}} \mathbf{T}\right\|^2_2.
			\label{constrainT}
		\end{aligned}
		$$
		As previously discussed, the parameter $\tau_2\geq0$ governs the influence of the constraints derived from the analyzer's time spectrum. A larger value of $\tau_2$ enforces a stronger constraint, meaning the quantity $\left|\mathbf{F}_{:, k}^{\mathrm{H}} \mathbf{C}_{\mathbf{T}} \mathbf{T}^{(m)}\right|^2$ is compelled to approximate the measured spectrum $I_{\mathbf{T}}(k)$.
		
		Provided that the matrix
		$$\sum_{l=1}^L \mathbf{C}_l^{\mathrm{H}} \operatorname{diag}\left(\left|\mathbf{R}^{(m+1)}\right|^2\right)_{n \times n} \mathbf{C}_l+\tau_2 \mathbf{C}_{\mathbf{T}}^{\mathrm{H}} \mathbf{C}_{\mathbf{T}}$$ 
		is invertible, the estimator for the analyzer admits the following closed-form solution:
		
		$$
		\begin{aligned}
			\mathbf{T}^{(m+1)} &= \left( \sum_{l=1}^L \mathbf{C}_l^{\mathrm{H}} \operatorname{diag}\left(\left|\mathbf{R}^{(m+1)}\right|^2\right)_{n \times n} \mathbf{C}_l \right. \\
			& \quad + \tau_2 \mathbf{C}_{\mathbf{T}}^{\mathrm{H}} \mathbf{C}_{\mathbf{T}} \Big)^{-1} \\
			& \quad \times \left( \sum_{l=1}^L \mathbf{C}_l^{\mathrm{H}} \left( \overline{\mathbf{R}^{(m+1)}} \odot \mathbf{\Psi}_l^{(m+1)} \right) \right. \\
			& \quad \left. + \tau_2 \mathbf{C}_{\mathbf{T}}^{\mathrm{H}} \mathbf{\Psi}_{\mathbf{T}}^{(m+1)} \right).
		\end{aligned}
		$$
		\begin{figure*}[ht]
			\setlength{\abovecaptionskip}{-0.2cm} 
			\centerline{\includegraphics[width=1.00\linewidth]{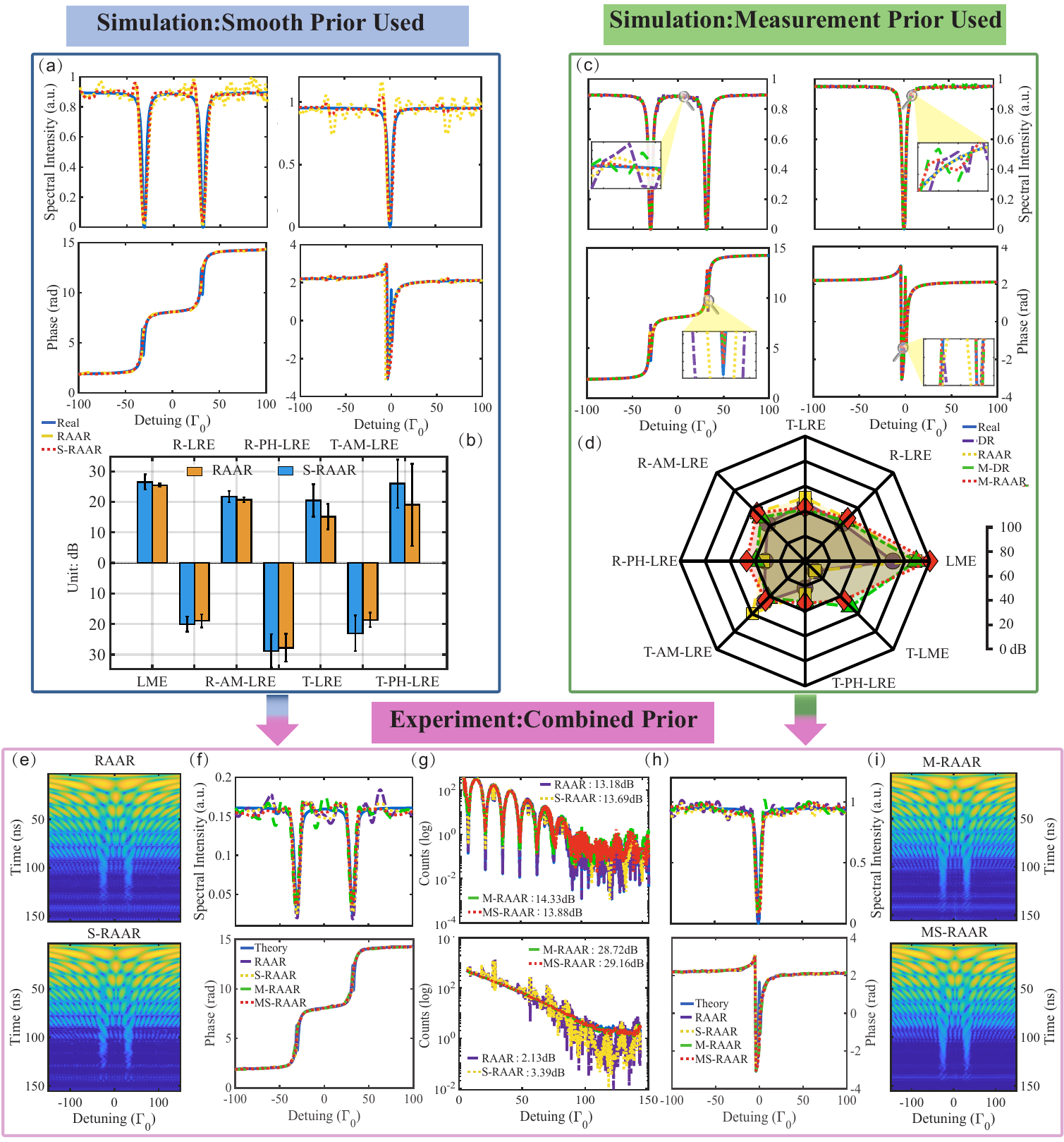}}
			\centering
			\begin{picture}(10,5)
			\end{picture}
			\caption{Performance evaluation of constrained optimization methods for blind ptychoscopy with an unknown analyzer. (a, b) Results obtained using a smoothness prior of the target response. (a) Recovered intensity and phase of $\mathbf{R}$ (left column) and $\mathbf{T}$ (right column) using the RAAR and S-RAAR methods. (b) Quantitative comparison of different methods via histograms of the following metrics: logarithmic measurement error (LME), logarithmic relative error of $\mathbf{R}$ (R-LRE), logarithmic relative error of $|\mathbf{R}|^2$ (R-AM-LRE), logarithmic relative error of $\text{phase}(\mathbf{R})$ (R-PH-LRE), logarithmic relative error of $\mathbf{T}$ (T-LRE), logarithmic relative error of $|\mathbf{T}|^2$ (T-AM-LRE), logarithmic relative error of $\text{phase}(\mathbf{T})$ (T-PH-LRE), and logarithmic of measurement error of time spectrum of $\mathbf{T}$ (T-LME). (c, d) Results obtained using a constrained time spectrum of the analyzer. (c) Intensity and phase profiles recovered by the DR, RAAR, M-DR, and M-RAAR methods. (d) Radar chart providing a multi-index comparison of the aforementioned methods across seven evaluation metrics. (e-i) Experimental results reconstructed by the RAAR, S-RAAR , M-RAAR, and MS-RAAR methods. (e, i) Recovered 2D spectra. (f, h) Recovered intensity and phase. (g) Comparison of time spectrum calculated from the recovered amplitude and phase for $\mathbf{R}$ and $\mathbf{T}$, alongside the experimental data. The R-LME and T-LME of the recovered results by each methods are also calculated.}
			\label{Fig.6} 
		\end{figure*}
		
		The evaluation of the proposed methods was conducted in two phases. First, the smoothness and measurement priors are tested individually on simulated data. Subsequently, they are combined to process experimental data. To comprehensively evaluate the performance of each method, nine different metrics are chosen: 
		\begin{itemize}
			\item Logarithmic measurement error (LME)
			\item Logarithmic relative error of $\mathbf{R}$ (R-LRE)
			\item Logarithmic relative error of $|\mathbf{R}|^2$ (R-AM-LRE)
			\item  Logarithmic relative error of $\text{phase}(\mathbf{R})$ (R-PH-LRE)
			\item Logarithmic relative error of $\mathbf{T}$ (T-LRE)
			\item Logarithmic relative error of $|\mathbf{T}|^2$ (T-AM-LRE)
			\item Logarithmic relative error of $\text{phase}(\mathbf{T})$ (T-PH-LRE)
			\item Logarithmic measurement error of time spectrum of $\mathbf{R}$ (R-LME) 
			\item Logarithmic measurement error of time spectrum of $\mathbf{T}$ (T-LME) 
		\end{itemize}
		These metrics are strategically designed to address the core characteristics of constrained blind Nuclear Ptychoscopy where both $\mathbf{R}$ (target) and $\mathbf{T}$ (analyzer) are unknown, and priors (smoothness and time-spectrum constraint) are integrated—covering three key evaluation dimensions to avoid one perspective from single metrics. Specifically:
		\begin{itemize}
			\item \textbf{Measurement fidelity}: LME quantifies the consistency between the reconstructed 2D time-energy spectrum (the core experimental data) and the noiseless ground truth, while R-LME/T-LME focus on the target or analyzer’s independent time-spectrum—directly verifying whether the recovered theoretical time-spectrum of target or analyzer fit the measured data.
			\item \textbf{Global reconstruction accuracy}: R-LRE and T-LRE assess the overall discrepancy between the reconstructed $\mathbf{R}/\mathbf{T}$ and their ground truths in the region of interest, reflecting whether priors mitigate the illness inherent in blind reconstruction.
			\item \textbf{Physical quantity-specific error}: R-AM-LRE/R-PH-LRE and T-AM-LRE/T-PH-LRE decompose errors into intensity and phase components—critical for Nuclear Ptychoscopy, as intensity corresponds to target density/analyzer response strength, and phase relates to refractive index/delay effects. These metrics pinpoint which physical quantity benefits more from the priors (e.g., whether smoothness primarily improves phase stability).
		\end{itemize}
		
		Consistent with the definition of LRE in Eq.~\eqref{LRE}, we also adopt logarithmic forms for other error metrics to facilitate visualization of reconstruction performance, as illustrated in the histograms and radar plots in Fig.~\ref{Fig.6}(a) and (d). In particular, the logarithmic measurement error (LME) quantifies the discrepancy between the recovered 2D spectrum $\tilde{\mathbf{I}}$ and the noiseless measurement $\mathbf{I}$, formulated as
		\begin{eqnarray}
			\text{LME}&=&-20\log_{10}\left(\text{Measurement Error}\right)\nonumber\\
			&=&20\log_{10}\frac{\|\mathbf{I}\|_2}{\|\tilde{\mathbf{I}}-\mathbf{I}\|_2}.\label{MRE}
		\end{eqnarray}
		The other logarithmic metrics are defined analogously, following the structure of Eqs.~\eqref{LRE} and \eqref{MRE}.

		The results, summarized in Fig.~\ref{Fig.6}, demonstrate the advantages of methods that incorporate prior information—specifically the smooth prior, measurement prior, and their combination—when addressing blind Nuclear Ptychoscopy. In particular, the smoothness-constrained RAAR (S-RAAR) method is evaluated across $20$ independent trials using simulated measurements at a SNR of $30$ dB. Compared to the standard RAAR algorithm, S-RAAR yields reconstructions with markedly reduced oscillations, especially in the recovered intensities of $\mathbf{R}$ and $\mathbf{T}$ displayed in Fig.~\ref{Fig.6}(a). This qualitative assessment is corroborated by the histogram of logarithmic measurement and relative errors presented in Fig.~\ref{Fig.6}(b), which indicates that the S-RAAR method achieved superior performance across all quantitative metrics.
		
		Subsequently, by leveraging the time-domain measurement of $\mathbf{T}$, the performance of the DR, RAAR, M-DR (DR with measurement prior), and M-RAAR (RAAR with measurement prior) methods are evaluated. As evidenced in Fig.~\ref{Fig.6}(c), the M-RAAR method produces the most accurate and stable results, exhibiting a notably improved reconstruction of the phase profiles for both $\mathbf{R}$ and $\mathbf{T}$ compared to the results in Fig.~\ref{Fig.4}(h). These findings highlight the substantial advantage of incorporating measurement constraints for solving the blind Nuclear Ptychoscopy problem when the response function of the analyzer is unknown. The quantitative results in Fig.~\ref{Fig.6}(d) further confirm that the M-RAAR method achieved the best values in five kinds of metrics, particularly in R-LRE and R-AM-LRE, which are essential for assessing the quality of the recovered spectrum.
		
		Finally, we apply the methods to experimental data, comparing RAAR, S-RAAR, M-RAAR, and MS-RAAR (RAAR with combined measurement and smooth priors). All methods produce results consistent with theoretical expectations for the 2D time–energy spectrum, as shown in Fig.~\ref{Fig.6}(e) and (i). The recovered intensities and phases, shown in Fig.~\ref{Fig.6}(f) and (h), also exhibit good agreement with theory. Importantly, the incorporation of a smooth prior in S-RAAR and MS-RAAR suppresses oscillations in the recovered $\mathbf{R}$ and $\mathbf{T}$, and this is consistent with the TV regularizer’s core mechanism of penalizing abrupt intensity variations. Furthermore, Fig.~\ref{Fig.6}(g) demonstrates that MS-RAAR provides the best overall agreement with the time-domain measurement of $\mathbf{T}$, achieving the highest T-LME value. Although M-RAAR shows marginally better performance on an isolated metric namely R-LME, the MS-RAAR reconstruction is characterized by substantially reduced oscillations in the magnitudes $|\mathbf{R}|^2$ and $|\mathbf{T}|^2$. Taken together, these results demonstrate that the combined use of smoothness and measurement priors provides the most robust and accurate reconstructions when the analyzer response function is unknown.
		
		\section{Conclusion}
		In this work, we present Nuclear Ptychoscopy, a versatile framework for reconstructing both the amplitude and phase of nuclear responses, even under partial prior knowledge or unknown analyzer conditions, validated through experimental data and simulations. This capability opens new opportunities for high-precision nuclear spectroscopy, such as accurately characterizing nuclear clock transition frequencies. The approach may be extended to imaging and spectroscopic studies in condensed matter, materials science, and chemical analysis, where phase-sensitive nuclear responses provide information complementary to conventional X-ray techniques. Furthermore, its compatibility with existing and emerging X-ray platforms—such as synchrotrons, XFELs, and XFEL oscillators~\cite{adams2019scientific,huang2023ming}—offers a practical approach accessible to a broad user community, enabling widespread application in X-ray and nuclear metrology.
		\section{Acknowledgment}
		\begin{acknowledgements}
			This work is supported by the National Key Research and Development Program of China under Contract No. 2024YFA1610900, the National Natural Science Foundation of China (NSFC) under Grant No.12501596, No.12471401, No.12475122, No.12447106, No.12147101 and No.62301089, and the Fundamental Research Funds for Provincial universities of Zhejiang under Contract No. GK249909299001-013. We acknowledge the support of the nuclear resonant scattering beamline of SPring-8 in Japan for Proposals No. 2020A1101 and 2021B1522.
		\end{acknowledgements}

		\bibliography{texref}
		
	\end{document}